\documentclass[12pt,a4paper,twoside]{book}
\usepackage[italian,english]{babel}
\usepackage[T1]{fontenc}
\usepackage[utf8]{inputenc}
\usepackage[font=small,labelfont=bf]{caption}
\usepackage{fancyhdr}
\usepackage{xspace}
\usepackage{amsmath}
\usepackage{ifthen} 
\newboolean{uprightparticles}
\setboolean{uprightparticles}{false} 
\usepackage{upgreek}
\usepackage{cancel}
\usepackage{slashed}
\usepackage{mathtools}
\usepackage{graphicx}
\usepackage{siunitx}
\usepackage{caption}
\usepackage{subcaption}
\usepackage{multirow}
\usepackage{tabularx}
\fancypagestyle{plain}{%
                       \fancyhead{}
                       
                      }
\numberwithin{equation}{chapter}
\begin{document}
	\pagenumbering{roman}          
	
	\pagestyle{empty}
	\begin{titlepage}
	\begin{center}
		{\LARGE{University of Ferrara}} \\[1ex]
	   	{\large{Physics and Earth Sciences Department}} \\[1ex]
	  	{\large{Master's Degree in Physics}} \\
		\vspace{0.7cm}
		\includegraphics[height=6cm]{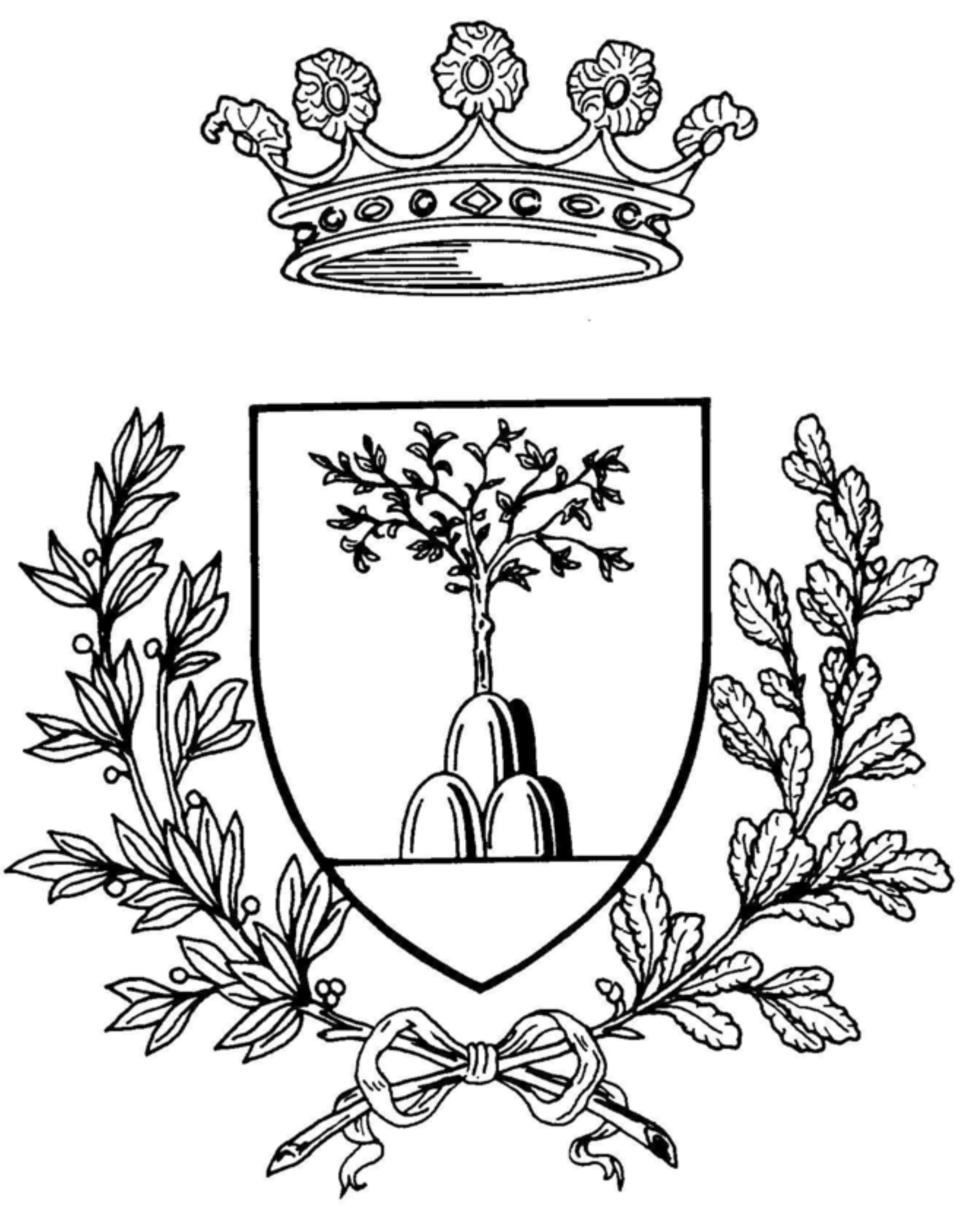}\\[1.5cm] 					
		{\Huge \textbf A cylindrical GEM detector for BES III}\\[1.5cm]
	\end{center}
	\begin{flushleft}
		{\em Supervisor}:  \\
		{\em Dott}. {\sc Diego Bettoni }\\[1.5ex]
		{\em Additional Supervisor}:  \\
		{\em Dott}. {\sc Gianluigi Cibinetto} \\[3.5ex]
		{\em Examiner}:  \\
		{\em Dott}. {\sc Massimiliano Fiorini } \\[3.5ex]
	\end{flushleft}
	\begin{flushright}
		{\em Candidate} \\
		{\sc Riccardo Farinelli} 
	\end{flushright}
	\vfill
	\begin{center}
		{ Academic year 2013-2014}
	\end{center}
\end{titlepage}

	\pagestyle{fancy}
	\markboth{Contents}{Contents}
	\baselineskip 5.5mm   
	\tableofcontents
	\baselineskip 5.8mm    
	\markboth{Contents}{Contents}
	\newpage
	
	\listoffigures
	\listoftables
	
	\cleardoublepage
	\pagenumbering{arabic}

	\chapter*{Introduction}

BESIII is a particle physics experiment located at the Institute of High-Energy Physics (BEPC-II) $e^+e^-$ collider at IHEP in Beijing. It takes data in the $\tau-charm$ domain since 2009. 
Currently, the world largest samples of $J/\psi$, $\psi(3686)$, $\psi(3770)$ and $\psi(4040)$ data have been collected. Among the many experimental results published so far, the large statistics accumulated at the $Y(4260)$ and $Y(4360)$ center of mass energies, allowed the discovery of the 
charged states $Z_c(3900)^+$ and $Z_c(4020)^-$ that are the first four-quark states observed and confirmed by different experiments.
BESIII data taking will last until at least 2022. \\

\noindent
The Italian collaboration is leading the effort for the development of a cylindrical GEM (CGEM) detector with analog readout to upgrade the current inner 
drift chamber that is suffering early ageing due to the increase of the machine luminosity. That will happen in either 2017 or in 2018.
The new detector will match the requirements for momentum resolution ($\sigma(p_t)/p_t \sim 0.5\%$ at 1 GeV) and radial resolution ($\sigma(xy) \sim100~\mu m$) 
of the existing drift chamber 
and will improve significantly the spatial resolution along the beam direction ($\sigma(z) \sim 150 \mu m$) with very small material budget (about 1\% of $X_0$). \\
The project, that now involves also groups from Mainz, Uppsala and IHEP, has been recognised as a Significant Research Project within the Executive Programme for Scientific and Technological Cooperation between Italy and P.R.C. for the years 2013-2015, and more recently it has been selected as one of the projects funded by the European Commission within the call H2020-MSCA-RISE-2014. \\

\noindent
Within the CGEM project, this work aims to perform full detector simulation for the optimisation of the tracker geometry and its operational
parameters. The goal is achieved by means of three different, but well connected, studies: a background estimation, a simulation of the detection elements and the
data analysis of a beam test. The thesis consists by five chapters.

\begin{itemize}
\item {\bf Chapter 1} describes the BESIII experiment, reporting about the physics program, the accelerator and the detector. 

\item {\bf Chapter 2} gives an overview of the CGEM project: the state of the art of the Micro Pattern Gaseous Detectors is presented, together with the innovations
and the design of the new BESIII inner tracker. At the end of this chapter a description of the cathode construction assembled in Ferrara is given.

\item {\bf Chapter 3} reports a study of the background estimation for the CGEM inner tracker: using a combination of real data and Monte Carlo simulation we extract 
noise information that is fundamental for the electronics design and to study the expected performance of the new detector.

\item {\bf Chapter  4} describes detailed simulation of the GEM detection element needed to provide information for the hit digitization in the Full Simulation. 
The parameters ({\it e.g.} size, shape, the Lorentz angle) of the electron avalanche are extracted for different gas mixture and operational parameters.

\item A beam test has been performed at CERN, at the end of 2014, in order to measure the performance of a BESIII GEM prototype in a magnetic field up to 1 Tesla.
The beam test results will be used also to validate the Garfield simulation and to provide additional information for the design of the detector and electronics. 
The first results of the beam test data analysis are reported in {\bf Chapter 5}.

\noindent
My contribution to the work reported in this thesis was broad and continuos during the past 12 months.
I participated to the construction of the cathode electrode, that was produced in Ferrara, helping during the assembling and
manufacturing procedures. For the background studies and the detector simulation I took care both of the framework development and
of the data analysis. Finally, I participated to installation and data taking of the beam test at CERN, and I was involved in the production of the reconstruction
and analysis software. During and after the beam test I participated to the processing and analysis of the data.

\end{itemize}

\chapter{The BESIII experiment}
\label{chBESIIIexp}
“Physis” is a Greek theological, philosophical and scientific term usually traslated into English as “nature”. Over the centuries scientists discovered the laws to describe an increasing number of physical phenomena from mechanics to electromagnetism, from the largest object in the universe to the tiniest event as the particles interaction.
High energy physics is a branch of physics which studies the nature of particles that are the constituents of what is usually referred to as matter and radiation. The current set of laws, concerning the electromagnetic, weak and strong interactions, forms the so-called Standard Model (SM). The Bejing Electron Spectrometer III (BESIII), located at the Bejing Electron Positron Collider II (BEPCII), is a particle physics experiment which aims to shed light on the nature of particle interactions within and beyond the SM.
 
The physics program \cite{BesIIIphysicsbook} of the experiment ranges from hadron spectroscopy to charm and charmonium physics, from $\tau$ physics to electroweak (EW) precision measurements in the energy range between 2 and 5 GeV.
These topics are described in the next paragraph.

\section{The Physics program}
\label{sec:physics}

The fundamental theory of the strong interactions, Quantum ChromoDynamics or QCD,  is well tested at short distances, but at long distances nonperturbative effects become important and these are not well understood. These effects are very basic to the field of particle physics and include $e.g.$, the structure of hadrons and the spectrum of hadronic states. Lower energy facilities with high luminosity can address these questions. Among these, the Beijing Electron Positron Collider II (BEPCII), which operates in the 2 GeV to 4.6 GeV energy range, plays a key role. This is because it spans the energy range where both short-distance and long-distance effects can be probed.\\

Theoretical studies of physics at the energy scale accessible to BEPCII continue to be actively pursued. To provide a good understanding of physics at this scale, theoretical tools derived from QCD have been invented. Since a full theoretical discussion is beyond the scope of this thesis I will just list the main theories and models used to describe the strong interaction in the non-perturbative regime. For charmonia, one can use the nonrelativistic QCD (NRQCR) and potential nonrelativistic QCD (pNRQCD) models to make theoretical predictions for physics involving both short- and long-distance effects, where a factorization of the two different kinds of effects can be accomplished and predictions that do not depend on the assumptions of any particular model can be made. For charmed hadrons, one can at least partly rely on heavy quark effective theory (HQET) for their study. For physics involving long-distance effects only, one can employ QCD sum-rule methods, or lattice QCD and make predictions from first principles. It is a fortunate coincidence that the most powerful tool for the quantitative treatment of nonperturbative dynamics, namely lattice QCD, is reaching a new level of sophistication with uncertainties in calculations of charmed quark dynamics that are approaching the 1 percent level. In addition to the theoretical tools derived from QCD, many phenomenological models have been invented to deal with nonperturbative effects, especially those at the 1 GeV scale or lower, such as light hadron spectroscopy, decays of charmonia and D-mesons into light hadrons, etc. Many theoretical predictions obtained with the above-mentioned methods
exist and call for tests from experiment. The BEPCII/BES-III facility is ideal for carrying out the task of confirming and validating these approaches.\\

Main topics, such as Charmonium spectroscopy, light-quark spectroscopy, D-physics and $\tau$ physic, are summarized in the following.

\subsection{Charmonium spectroscopy}

The total decay widths of the $J/\Psi$ and $\Psi$' are measured at a precision level that is
better than 1\%. The $J/\Psi$ has many different decay modes. In two-body decays, either of
the final-state particles can be a pseudoscalar, a scalar, a vector, an axial vector or a tensor
meson. With a 10$^{10}$ J/$\Psi$ event sample, these decay modes can be measured much more
precisely than before. Historically, there are some notorious problems related to decays of
charmonia. Among them the most well known problems are the $\rho\pi$ puzzle, $i.e.$ violations
of the 12\%-rule, and non-$D-\overline{D}$
decays of the $\Psi$(3770). With BESIII’s huge data samples,
more detailed experimental information will be forthcoming that will hopefully provide
guidance leading to solutions of these problems. Transitions between various charmonium
states will be measured with unprecedented precision. With the possibility of running at
higher energies, the recently discovered $Y(4260)$ \cite{4260} has been accessed at BEPCII, and this
led to the discovery of  $Z_c^+(3900)$ \cite{3900} the first observation of a tetra-quark or a charm molecule confirmed later by other experiments, and the $Z_c(4020)$.
With such huge data samples, it is possible to detect some Cabbibo-suppressed
$J/\Psi$ decay channels. In these channels, the charmed quark decays via the weak interaction,
while the anticharm quark combines with another quark to form a D-meson. This process
will provide the possibility for detecting effects of new physics at BEPCII, if, for example,
branching ratios of those decays are found to be larger than SM predictions. Also, one
can search for evidence of flavor-changing neutral currents. This an area where BESIII
can make unique explorations for physics beyond the SM.

\subsection{Light-quark spectroscopy}

Using $J/\Psi$-decays, one can study light hadron spectroscopy and search for new hadronic
states. The large $J/\Psi$ sample makes BEPCII a “glue” factory, since the charmed- and anticharmed quark constituents of the $J/\Psi$ almost always annihilate into gluons. This is very
useful for glueball searches and for probing the gluon contents of light hadrons, including
the low-lying scalar mesons.
QCD predicts the existence of glueballs and lattice QCD predicts their masses. For
example, the $0^{++}$ glueball is predicted to have a mass that is between 1.5 and 1.7 GeV. But
to date the existence of these various glueballs has still not been experimentally confirmed.
Also, since QCD is a relativistic quantum field theory, any hadron should have some gluon
content if symmetries allow. These gluon contents, especially those in scalar mesons, are
crucial inputs to the understanding of the properties of the light hadrons, such as the
$f_0(1500)$ and $f_0(1700)$ scalar mesons. The rich gluon environment in $J/\Psi$ decays is an ideal place
to study these issues.
Recently, evidence for exotic hadrons, $i.e.$ mesons that cannot be classified as a $q\overline{q}$ state
of the traditional quark model, have been seen experimentally. In principle, QCD allows
for the existence of exotic hadrons. With high-statistics data samples, comprehensive
searches for exotic states can be performed and the quantum numbers of any candidates
that are found can be determined.
In BESIII, an anomalous near-threshold mass enhancement is seen in the $p\overline{p}$ system
produced in the radiative decay process $J/\Psi \rightarrow\gamma p\overline{p}$; similar enhancements are seen in
other baryonic systems. Various explanations for these enhancements have been proposed,
$e.g.$, there may be resonances just below the mass thresholds. However, a satisfactory and
conclusive explanation has still not emerged. 

\subsection{$D$ physics}

At BEPCII, $D^+$ and $D^0$ mesons are produced through the decays of the $\Psi(3770)$,
and $D_s$ mesons can be produced through $e^+e^-$ annihilation at $\sqrt{\mathcal{s}}$ around (4.03 GeV). 
The decay constants $f_D$ and $f_{D_s}$ can be measured from purely leptonic decays with a
systematic errors of 1.2\% and 2.1\%, respectively. Inclusive and exclusive semileptonic
decays of $D$-mesons are also studied to test various theoretical predictions. Moreover,
through the study of the decays $D^0\rightarrow K^- e^+ \nu_e$ and $D^0\rightarrow \pi^- e^+ \nu_e$ one can extract the
CKM matrix elements V$_{cs}$ and V$_{cd}$ with a systematic error of around 1.6\%.
With BESIII it is possible to measure $D-\overline{D}$ mixing and search for $CP$-violation.
Theoretical predictions for mixing and $CP$-violation are unreliable; BESIII can provide
new experimental information about them.
Rare- or forbidden decays can provide strict tests of the SM and have the potential
of uncovering the effects of new physics beyond the SM. With BESIII, they can be
studied systematically. 

\subsection{$\tau$ physics}

$\tau$-physics will also be studied at BESIII, where several important measurements can be made. 
Experimental studies of inclusive hadronic decays can provide precise determinations 
of the strange quark mass and the CKM matrix element V$_us$, while the study of
leptonic decays can test the universality of the electroweak interaction and give a possible hint of new physics.


\section{BEPC II}

\begin{figure}[btp]
		\centering
   		\includegraphics[width=0.8\textwidth]{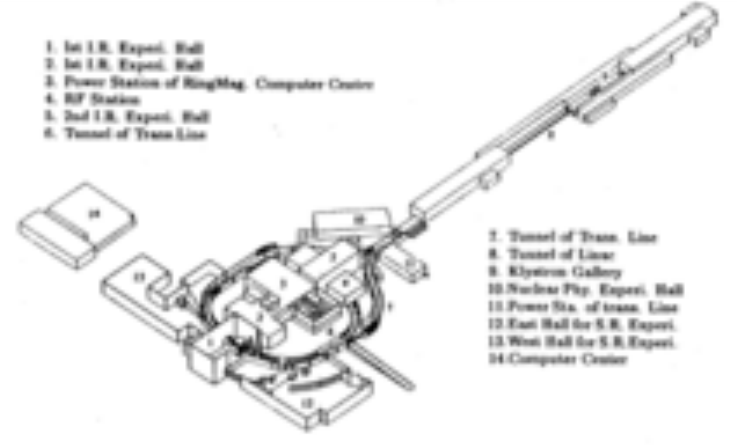}
   		\caption[BEPCII layout.]{A layout of the BEPCII facility; the LINAC and $e^+e^-$ storage ring are shown.}
	\label{fig:BEPCII}
	\end{figure}

\begin{figure}[btp]
		\centering
   		\includegraphics[width=0.8\textwidth]{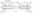}
   		\caption[Beampipe cross section]{Beampipe cross section and its mechanical support.}
	\label{fig:beampipe}
	\end{figure}

The BEPCII-accelerator consists of a 202 m long electron-positron LINAC injector and a double-ring structure as shown in fig \ref{fig:BEPCII}. The inner ring and the outer ring cross each other in the northern and southern interaction points (IP). The horizontal crossing angle between two beams at the southern IP, where the BESIII detector is located, is 11 mrad to meet the requirement of sufficient separation but no significant degradation to the luminosity. While in the northern crossing region, the two beams cross horizontally and a vertical bump is used to separate two beams, so that the optics of the two rings can be symmetric. Fig \ref{fig:beampipe} shows a sketch of the beam line. For the dedicated synchrotron radiation operation of the BEPCII, electron beams circulate in the outer ring with a pair of horizontal bending coils in super conducing (SC) magnets serving this purpose and in the northern IP a bypass is designed to connect two halves of the outer ring \cite{CZhang}.\\

The BEPC injector is a 202-meter electron/positron LINAC with 16 RF power sources and 56 S-band RF structures. The BEPCII main requirements for the injector are: the full energy of $e^+$ and $e^-$ beams injected into the storage rings, $e.g.$ $E_{inj} \geq $ 1.89 GeV and the $e^+$ injection rate $\geq$ 50mA/min.


Luminosity is one of the most important parameters, in a $e^+e^-$ collider is expressed as:

	\begin{equation}
	\label{eq:Lumino}
		\mathcal{L}(cm^{-2}s^{-1}) = 2.17 * 10^{34}(1+r)\xi_y \frac{E(GeV)k_bI_b(A)}{\beta_y^*(cm)}
	\end{equation}

Where r = $\sigma^*_y / \sigma^*_x$ is the beam aspect ratio at the IP, $\xi$ the vertical beam-beam parameter, $\beta^*_y$ the vertical envelope function at IP, $k_b$ the bunch number in each beam and $I_b$ the bunch current. With the BEPCII parameters the luminosity can reach $8.04 \times 10^{32} $cm$^{-2} s^{-1}$ \cite{lumi}.
Tab. \ref{table:BEPCII} summarises the main parameters.

\begin{table}[ht]
\begin{minipage}{\textwidth}
\begin{center}
\begin{tabular}{lcc}
\hline
Beam Energy E			& GeV	&	1.55(1.89)	\\
\hline
Circumference			& m	&	237.53		\\
\hline
Bunch Number $k_b$		& 	&	93		\\
\hline
Beam Currents per Ring $I_{beam}$& mA	&	1116		\\
\hline
RF Frequency $f_{RF}$		& MHz	&	499.8		\\
\hline
RF Voltage per ring $V_{RF}$	& MV	&	1.5		\\
\hline
Bunch Lenght			& cm	&	1.1/1.5		\\
\hline
Bunch Spacing			& m 	&	2.4		\\
\hline
Beam-BeamParameter $\beta_x\beta_y$&  	&	0.04/0.04	\\
\hline
Crossing angle			& mrad	&	$11^2$		\\
\hline
Luminosity			& $cm^{-2}s{-1}$& $1 \times 10^{33}$	\\
\hline
\end{tabular}
\caption[BEPCII parameters.]{BEPCII parameters.}
\label{table:BEPCII}
\end{center}
\end{minipage}
\end{table}

The SC RF cavities are chosen for their advantage of large accelerating gradient and well-damped HOMs. Two SC cavities are installed in the BEPCII with one cavity in each ring to provide necessary RF voltage of 1.5 MV. Each cavity is powered with a 250 kW klystron. The horizontal high power test gives the Q values of $5.4 \times 10^8$ and $9.6 \times 10^8$ at $V_rf$=2 MV for the west and east cavities, higher than the design values of $5 \times 10^8$ at 2 MV.

The BEPCII re-use 44 BEPC bends and 28 quads. There are 267 new magnets, including 48 bends, 89 quads, 72 sextupoles, 4 skew quads and 54 dipole correctors.

In order to meet the challenges both on the filed uniformity and low coupling impedance, a modified slotted pipe kicker has been developed with the coating strips on ceramic bar instead of metallic plates as the beam image current return paths.
The vacuum system has to ensure a vacuum pressure of $8 \times 10^{-9}$ Torr.
The interaction region (IR) has to accommodate competing and conflicting requirements from the accelerator and the detector. Many types of equipment including magnets, beam diagnostic instruments, masks, vacuum pumps, and the BESIII detector must co-exist in a crowded space. A special pair of superconducting magnets (SCQ) is placed in the IR. Each SCQ consists of a main and a skew quadrupoles, 3 compensation solenoids and a dipole coils, to squeeze the $\beta$  function at IP, compensating the detector solenoid and to serve as the bridge connecting outer ring for SR operation, respectively.
The BEPCII cryogenics system is composed of four sub-systems: the central cryogenic plant and three satellite cryogenic systems for the RF cavities, the SCQ magnets, and the SSM detector solenoid. Two 500 W refrigerators serve the purpose to cool the SC devices at 4.5 K, one for the cavities and another for the magnets.

If BESIII is not running, the BEPCII rings are used for synchrotron radiation with a beam current of 250 mA at 2.5 GeV and 150 mA at 2.8 GeV respectively.

\section{The BES III detector}

BESIII is a multi-purpose detector built for the physics explained in Sec. \ref{sec:physics} with a set of sub-detectors: the multilayer drift chamber, time of flight detector, an electro-magnetic calorimeter, a super-conducing solenoid and a muon chamber. The polar angle coverage of BESIII goes from \ang{21} to \ang{159} and the solid angle coverage is $\Delta\Omega / 4\pi$ = 0.93 \cite{DesConsBESIII}. 

\begin{figure}[btp]
		\centering
   		\includegraphics[width=1.1\textwidth]{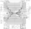}
   		\caption[Schematic drawing of BESIII detector.]{Schematic drawing of the BESIII detector; the barrel angular coverage is -83$^{\circ}$ < $\theta$ < 83$^{\circ}$ and the endcap is  $\pm$ 83$^{\circ}$ < $\theta$ < $\pm$ 93$^{\circ}$.}
	\label{fig:BESIII}
	\end{figure}

The BESIII detector, shown in Fig. \ref{fig:BESIII}, can be divided in two part: the barrel, with a cylindrical simmetry, covering the central region, and the endcap, in the forward and backward direction.

BESIII is built around a 1 T superconducting solenoid (SSM). The electromagnetic calorimeter is located inside the coil of the superconducting magnet which has a mean radius and length respectively of 1.482 and 3.52 m. The innermost detector is the multilayer drift chamber (MDC) that surrounds the beryllium beam pipe, then there is the time-of-flight (TOF) system consisting of two layers of plastic scintillator counters. The CsI(Tl) electromagnetic calorimeter (EMC) is placed between the TOF and the SSM. The muon identifier (MU) consist of layers of resistive plate chambers (RPCs) inserted in gaps between steel plates of the flux return joke. The main parameters and performance of the various detector sub-systems are shown in Tab. \ref{table:BESIII}.

\begin{table}[ht]
\begin{minipage}{\textwidth}
\begin{center}
\begin{tabular}{ccc}
\hline
\textbf{Sub-system}	& 				& \textbf{Value} 	\\
\hline
MDC	& Single wire resolution ($\mu$m)	& 130		\\
\hline
MDC	& $\sigma_p/p$ (1GeV/c)			& 0.5\%		\\
\hline
MDC	& $\sigma$ (dE/dx)			& 6\%		\\
\hline
TOF	& $\sigma_T$ (ps) barrel		& 100		\\
\hline
TOF	& $\sigma_T$ (ps) EndCap 		& 110		\\
\hline
EMC	& $\sigma_E/E$ (1GeV)			& 2.5\%		\\
\hline
EMC	& Position resolution (1GeV)		& 0.6 cm	\\
\hline
MU	& number of layers (barrel/endcap)	& 9/8		\\
\hline
MU 	& Cut-off momentum (MeV/c)		& 0.4		\\
\hline
SSM	& Solenoid magnet Filed (T)		& 1.0		\\
\hline
	& $\Delta\Omega/4\pi$			& 93.4\%	\\
\hline
\end{tabular}
\caption[Detector parameters and performance.]{Detector parameters and performance.}
\label{table:BESIII}
\end{center}
\end{minipage}
\end{table}

\subsection{The Multilayer Drift Chamber}
MDC \cite{BesIIIphysicsbook} is optimized to track the low momentum particles (below 1 GeV) with a good momentum resolution and dE/dx measurement capability with a resolution of $\sim$ 4.5\% for identifying charged particles. The main function is to reconstruct charged tracks in 3D space or short-lived hadrons as $K^0_s$ that decay in the MDC volume, to provide extrapolated track positions at the other detector components.

The MDC also produces signals for the level 1 triggers to select good physics events and reject the background. The inner radius is 59 mm and the outer is 810 mm. The gas-mixture is $He-C_3H_8$ 60:40 with water vapor, it is chosen to minimize the multiple scattering effect. The single cell position is around 130 $\mu$m in the $r-\phi$ plane and $\sim$ 2 mm in the z-coordinate. A 3$\sigma$ $\pi/K$ separation is achieved up to 770 MeV/c with a 6\% dE/dx resolution for particles with incident angle of \ang{90}.\\

\begin{figure}[btp]
		\centering
   		\includegraphics[width=0.8\textwidth]{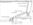}
   		\caption[The MDC mechanical structure.]{The MDC mechanical structure.}
	\label{fig:mdcmech}
	\end{figure}

In the MDC the cell consists of a sense wire surrounded by 8 field wires. The cell dimension is identical for each cell in different layers in order to keep the gas gain the same. The single wire resolution is dominated by electron diffusion. In order to minimize the material the field wires are made of 110 $\mu$m diameter gold plated aluminium and the sense wire are gold plated tungsten with 3\% rhenium. The drift chamber contains 43 sense wire layers arranged as 11 superlayers. Layers 1 to 8 and 21 to 36 have a small stereo angle. Layers 9 to 20 and 27 to 43 are axial. These layers are divided in inner chamber and outer chamber, as shown in Fig. \ref{fig:mdcmech}.

The measurements are obtained by stereo wire superlayer with stereo angles in the range of \ang{-3.4} to \ang{3.9}. 

In multulayer tracking chamber with homogeneously spaced wires layer along the particle trajectories in a uniform axial magnetic field, the following expression is used to estimate the transverse momentum resolution :

\begin{equation}
\label{eq:transmom}
	\frac{\sigma_{p_t}}{p_t} = \sqrt{\left( \frac{\sigma_{p_t}^{wire}}{p_t} \right) + \left( \frac{\sigma_{p_t}^{MS}}{p_t} \right)} 
\end{equation}

where $p_t$ is the transverse momentum, $\sigma_{p_t}^{wire}$ is the momentum resolution of the position measurements of individual wires and $\sigma_{p_t}^{MS}$ is the momentum resolution contribution due to multiple scattering of tracks inside the tracking chamber.
If we assume $p_t$ = 1 GeV/c and B = 1 T, then the expected momentum resolution at 90$^{\circ}$ is

\begin{equation}
\label{eq:transmom2}
	\sigma_{p_t} = \sqrt{ \sigma_{p_t}^{wire} + \sigma_{p_t}^{MS}} = \sqrt {0.32\%+0.35\%} = 0.47\%. 
\end{equation}

This result agrees with the Monte Carlo calculations \cite{MdcMC}.

The dE/dx performance is studied from the data from the seven outer layers. The deposited energy distribution has a Landau shape and it is truncated to convert it to a Gaussian-like spectrum and to reduce the tail fluctuation. A dE/dx resolution of 4.5\% is achieved.

The cross-sectional view of one quarter of the MDC mechanical structure is shown in Fig. \ref{fig:mdcmech}. The MDC consists of two parts, inner and outer chamber, joined together at the end plate. The end plates of the outer chamber have two different shapes, a conical section and a stepped section.

Over the last years the MDC efficiency has been decreasing due to ageing issues (see Sec. \ref{sec:MDCaging}).

\subsection{Time-of-flight (TOF)} 
TOF detector systems based on plastic scintillation counters have been very powerful tools for particle identification in collider detectors. In addition the TOF system can provide fast trigger signals. 

The system is composed by a double layer barrel plus two endcaps and it is placed between the MDC and the EMC. Each layer has 88 plastic scintillation counters. The endcap TOF counters are placed outside the two MDC end plates \cite{BesIIIphysicsbook}. 

The polar coverage is $|cos\theta| \leq$ 0.82 for the barrel and 0.85 $\leq |cos\theta| \leq 0.95$ for the endcaps. The scintillator length is 2300 mm. The bar cross section is trapezoidal and the thickness is 50 mm. Two PMTs are attached to the two ends of a barrel counter and coupled by 1mm thick silicone pads. 

The time resolution is determined by the rise time of the scintillation light, the fluctuations of  photon arrival time at the PMT and the transition time spread of the PMT. The average time resolution is 90 ps in the barrel and 120 ps in the endcap. A 3 $\sigma$ $K/\pi$ separation is a requirement of the TOF. 

To improve it a likelihood analysis is used. Combining TOF time resolution and MDC dE/dx measurement, the efficency of K/$\pi$ separation is about 95\%.

\subsection{Electro-Magnetic Calorimeter} 
The main requirement for the EMC \cite{BesIIIphysicsbook} is high energy and position resolution to give a good e/$\pi$ separation and to detect direct photons and those ones from the $\pi^0,\eta, \rho$ decays. The EMC allow also to separate direct photon to those from the decays. The average photons multiplicity is four per event, similar to the multiplicity of charged particles. The energy range of photons goes from 20 MeV to the GeV scale. 

The EMC is the only one detector that can detect photons and an important requirement is to reconstruct accurately the $\pi^0$ invariant mass in $\pi^0\rightarrow\gamma\gamma$, a process with an opening angle that decreases as the energy increases. At 1.5 GeV the opening angle is about \ang{10}.

Charged particles can interact with the calorimeter and generate showers. Every particles generate a special patterns due the energy deposited. At low energy there is a misidentification of electrons and pions. High segmentation allows to distinguish electron showers from hadron showers.

The calorimeter is placed inside the 1T SSM. The CsI(Tl) scintillating crystals are the best choice for the BESIII EMC that must detect low energy photons. Choices made for the CsI(Tl) calorimeter, including crystal size, length, segmentation and geometric parameters, are based on physics requirements described above. 

The crystal length is 28 cm and there are 6240 crystals arranged as 56 rings. Each crystal covers an angle of about \ang{3} in both polar and azimuth directions. The barrel and the two endcaps are separated by 5 cm gaps.

The light generated inside the crystals is readout by Hamamatsu S2744-08 photodiodes \cite{DesConsBESIII}.

\subsection{Superconducting Solenoid Magnet}
\label{sec:ssm}
The 1.0T SSM \cite{Zhu} allows precise momentum measurements of charged particles. The steel flux return is used as hadron absorber for hadron/muon separation and provides the overall structure and support of the BESIII components. 

The 0.7 mm diameter NbTi/Cu superconductor strands were formed into the 12 strand Rutherford cable. The total thickness of the coil ground insulator required to withstand 2000 V. 

Fifteen layers of super insulation films separate the liquid nitrogen thermal shield and the liquid helium cooled cold mass. The yoke steel is divided into nine layers and the thickness is chosen to optimize the performance of the muon identification. The material for the yoke is low carbon steel that has sufficient strength and acceptable magnetic properties.

\subsection{Muon Chamber}
Identification of muons is of great importance in BESIII physics. The muon detector needs high identification efficiency with an acceptance as large as possible and a momentum cut-off as low as possible. As shown in Sec. \ref{sec:ssm}, the muon chamber \cite{BesIIIphysicsbook} is built in the steel yoke of the SSM. By associating hits in muon counters with tracks reconstructed in the MDC and energy measured in the EMC, muons can be identified with low cut-off momentum.

The BESIII muon chamber is made of to resistive plate counters (RPCs) placed in the steel plates of the magnetic flux return and divided in nine layers. The RPC modules have readout strips in both the $\theta$ and $\phi$ directions for reconstructing charged particle tracks. The average width of the strips is about 4 cm.

\subsection{MDC ageing issues}
\label{sec:MDCaging}

Due to beam background, the inner chamber of the MDC shows ageing effect. The gain of the cells of these detector part is decreasing up to 25\% for the first layer \cite{1403}. 

\begin{figure}[btp]
		\centering
   		\includegraphics[width=0.8\textwidth]{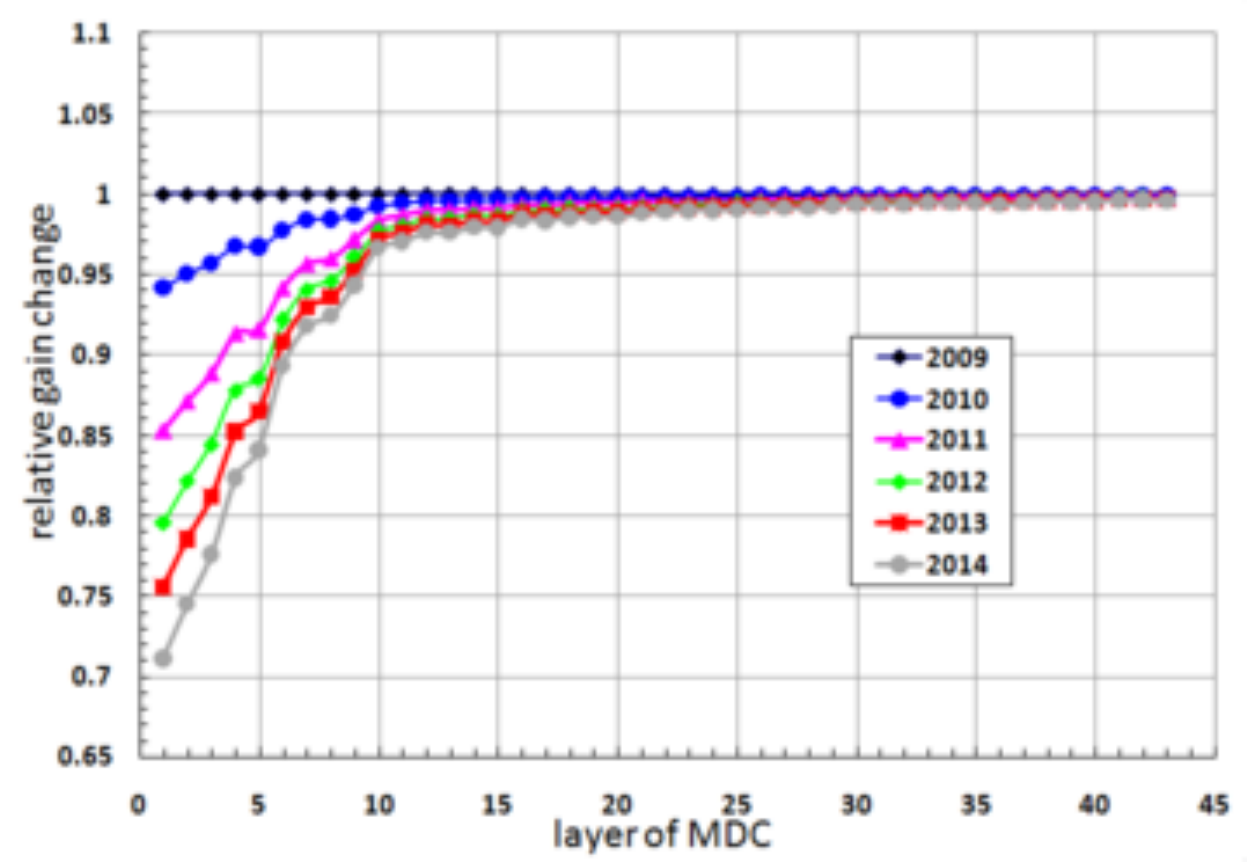}
   		\caption[MDC ageing effect in relative gain]{Relative gain change versus the layers number for the last 6 years. The MDC ageing effect is evindet: the first 15 layers have lost gradually their gain.}
	\label{fig:MDCeff}
	\end{figure}

In order to reduce the dark currents of the sense wires and to use the detector in a safe region the voltage of the first four layers has been decreased with respect to the normal values, which result in the decrease of the performance of the detector. This worsening does not allow the MDC to mantain the required detector performance in the next years.
Fig. \ref{fig:MDCeff} shows the behaviour of the gain for different years.

If the background remains at the level as the latest years, the gain of the MDC first layer will decrease about 4\% each year, so we can estimate that the gain of the first layer will become 63\% by 2016.

\begin{figure}[btp]
		\centering
   		\includegraphics[width=0.8\textwidth]{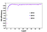}
   		\caption[MDC ageing effect in hit efficiency]{Hit efficiency versus the layer number for the past 3 years. A drop of efficiency is clear for the first 3 layers due the MDC ageing.}
	\label{fig:mdchiteff}
	\end{figure}

Fig. \ref{fig:mdchiteff} shows the hit efficiency versus the layer number. Due to high degradation of the first 3 layers the hit efficiency decrease significantly.

Being BESIII an experiment that has to take data at least 8 more years, the inner part of the MDC has to be changed with a new and more performant inner tracker\cite{CDR,1403}.

	\chapter{Cylindrical GEM based Inner Tracker}
\label{chCGEM}

A possible solution to the ageing problem of the inner chamber of the MDC would be to built a new Inner Tracker (IT) consisting of Cylindrical Gas Electron Multuplier (CGEM) deterctors. A Conceptual Design Report (CDR) \cite{CDR} with this proposal, presented at the BESIII Collaboration Meeting of June 2014 and then it has been accepted by the Executive Board and by the Institutional Board of the collaboration. 

\begin{figure}[btp]
		\centering
   		\includegraphics[width=0.8\textwidth]{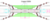}
   		\caption[Three CGEM layers layout.]{The green bars represent the new CGEM IT around a schematic drawing of the beampipe.}
	\label{fig:ciccia}
	\end{figure}

The CGEM IT consists of three indipendent tracking layers (L1-L3), each one has a 2-D readout and it provides a 3-D reconstruction, see Fig. \ref{fig:ciccia}. Each layer has a cylindrical shape to fit the required coverage around the beam pipe and it is made of a triple-GEM, a cathode and an anode plane, according to a design developed by the KLOE-2 experiment and shown in Fig. \ref{fig:triplegem}. 

The readout anode of each CGEM is segmented with 650 $\mu$m pitch $XV$ patterned strips with a stereo angle that changes depending on the layer geometry. The full system consists of about 10,000 electronics channels.

This technology allows the construction of a light and fully sensitive detector, that can match the stringent requirement on the material budget needed to minimize the multiple scattering effect for low-momentum tracks and improving the current vertex resolution of the apparatus.

An overview of the project will be presented in this chapter.

\section{Requirements}

The CGEM IT detector is required to feature significant radiation hardness, high rate capability, excellent spatial resolution both in the longitudinal and transverse directions, good time resolution and limited total radiation length. 

The space available in the inner part of the BESIII spectromenter is the one that will be left free by the removal of the inner part of the Drift Chamber (DC): it is limited and this fact introduces significant constraints on the mechanical design and on the power dissipation per readout channel. \\

A Gas Electron Multiplier (GEM) detector offers a technique to address such requirements as radiation hardness and high rate capability: furthermore the building technique allows a flexible geometry.

To match the available space and meet the BESIII physical requirements, the upgrade of the inner drift chamber should take into account the following points:
\begin{itemize}
\item the radial extension goes from 78 mm to 179 mm and the coverage is $0.93 \pi$;
\item the spatial resolution has to be better than 100 $\mu$m in the $r$-$\phi$ view and $\sim$ 200 $\mu$m in the $r$-$z$ view to be comparable with the resolution provided by the current DC-IT;\footnote{$z$ is the coordinate along the beam line, $r$ is the radial coordinate and $\phi$ the azimuthal angle.}
\item the material budget has to be less than 1.5\% of $X_0$;
\item the rate capability has to be larger than 4 kHz/cm$^2$;
\item the momentum resolution $\sigma_{P_t}/P_t$ $\sim$ 0.5 \% at 1 GeV/c;
\item the efficiency of the detector has to be of the order of or better than 98 \%.
\end{itemize}

\section{State of art}
\label{art}
Gas-filled detectors localize the ionization produced by charged particles, generally after charge multiplication. When an ionizing particle passes through the gas it creates electron-ion pairs. Multi-wire proportional and drift chambers were the first detectors to use this physical effect but these are overcome by micro-pattern gas detectors (MPGD) due their limitation by basic diffusion processes and space charge effects. Modern photolithograpic technology led to the development of MPGD concepts \cite{Sauli2}. The Micro-Strip Gas Chamber (MSGC), invented in 1988, was the first micro-structure gas chamber \cite{Oed}. It consists of a set of tiny parallel metal strips laid on a thin resistive support, alternatively connected as anodes and cathodes. Further studies led to develope more powerful devices such as GEM and Micromega \cite{ChinisePhysics}. These have improved reliability and radiation hardness. The absence of space charge effects has upgraded the maximum rate capability (> 10$^6$ Hz/mm$^2$) \cite{ChinisePhysics}. 

\begin{figure}[btp]
		\centering
		\begin{subfigure}[h]{0.4\textwidth}
   			\includegraphics[width=\textwidth]{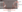}
			\caption{}
			\label{fig:gemhole:a}
		\end{subfigure}
		\begin{subfigure}[h]{0.4\textwidth}
   			\includegraphics[width=\textwidth]{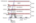}
			\caption{}
			\label{fig:gemhole:b}
		\end{subfigure}
   		\caption[GEM hole and triple GEM.]{a) Picture of the GEM hole cross section with biconical shape. Dimensions are shown in the figure. b) Triple GEM schematic drawing.}
	\label{fig:gemhole}
	\end{figure}

The GEM foil is made by 50 $\mu$m Kapton foil, copper clad on each side, with a high surface density of holes \cite{Sauli}. Each hole has a bi-conical structure with external (internal) diameter of 70 $\mu$m (50  $\mu$m); the hole pitch is 140 $\mu$m as shown in Fig. \ref{fig:gemhole:a}. A recent new photolithographic process, the so called single-mask technology \cite{singlemask}, has overcome the limitation given by the double-mask technology. The single-mask GEM sheets have a sheet width up to 550 $\mu$m.
The GEM foils are manufactured by the CERN EST-DEM workshop. \\

A typical voltage difference of 200 - 400 V is applied between the two copper sides, giving fields around 10$^2$ kV/cm. A good working GEM has a gain of the order of 100, thus then multiple structures realized by assembling two or more GEMs at close distance allow high gains to be reached while minimizing the discharge probability \cite{Bachmann}. 
The cathode and anode foil are made by 50 $\mu$m Kapton foil with a copper cladding of 3 $\mu$m on the internal surface and the cathode has a simple design because it is used to induce the high voltage; on the contrary the anode has a complex design due to the readout strips. The full anode design is described in Sec. \ref{sec:Innovations}.

Electrons released by primary ionization in the conversion region of the GEM drift into the holes, where charge multiplication occurs in the high electric field. Several GEM foils allow to operate at overall gas gain above 10$^4$. The typical layout of a triple GEM detector is shown in Fig. \ref{fig:gemhole:b}. Together with the three GEM foils, providing the amplification, the cathode and the anode complete the design; the former provides the drift field and the latter is segmented in strips and collects the charge. With this geometry four regions are defined: drift where the primary electrons are mainly produced, transfer 1 and 2 the regions between the GEMs plane and the induction region where the electrons are collected. The performance and robustness of GEM detectors motivate their use in high-energy and nuclear physics.\\

COMPASS \cite{Ketzer}, a high luminosity experiment at CERN, pioneered the use of large-area planar triple GEM ($\sim$ 40 x 40 cm$^2$) close to the beam line with a particle rate of 25 kHz/mm$^2$. The technology achieved a tracking efficiency close to 100\%, a spatial resolution of $\sim$ 70 $\mu$m and a time resolution of the order of ns. 
The GEM tracking detector for COMPASS shows that the triple GEM technology allows to reduce by more than one order of magnitude the probability of a gas discharge due by the fact that, for a given gain, the triple GEM can be operated at much lower voltages over each GEM foil so a gain of $\sim$ 8000, high value for a gas detector, can be achieved and have a better S/N ratio.
COMPASS is working with an analog readout digitized by a 10-bit ADC: on each channel the charge depositated is measured and used to weigh the signal deposited over the strips. GEM detectors are also used in LHCb as trigger in the Muon Chamber and as tracking in the TOTEM Telescope. 

The main difference between the COMPASS detector and the BESIII CGEM is the magnetic field: COMPASS achieved these results without magnetic field. The gas mixture choosen is Argon-CO$_2$ (70:30)\footnote{from now on, Argon-CO$_2$ (70:30) will be referred as Argon-CO$_2$} because it allows low diffusion and large drift velocity of electrons and non-polymerization in case of discharge. The gaps configuration is 3/2/2/2 (drift/trasfer1/transfer2/induction) and a strip pitch of 400 $\mu$m in $XY$ view.\\

	\begin{figure}[btp]
		\centering
   		\includegraphics[width=0.8\textwidth]{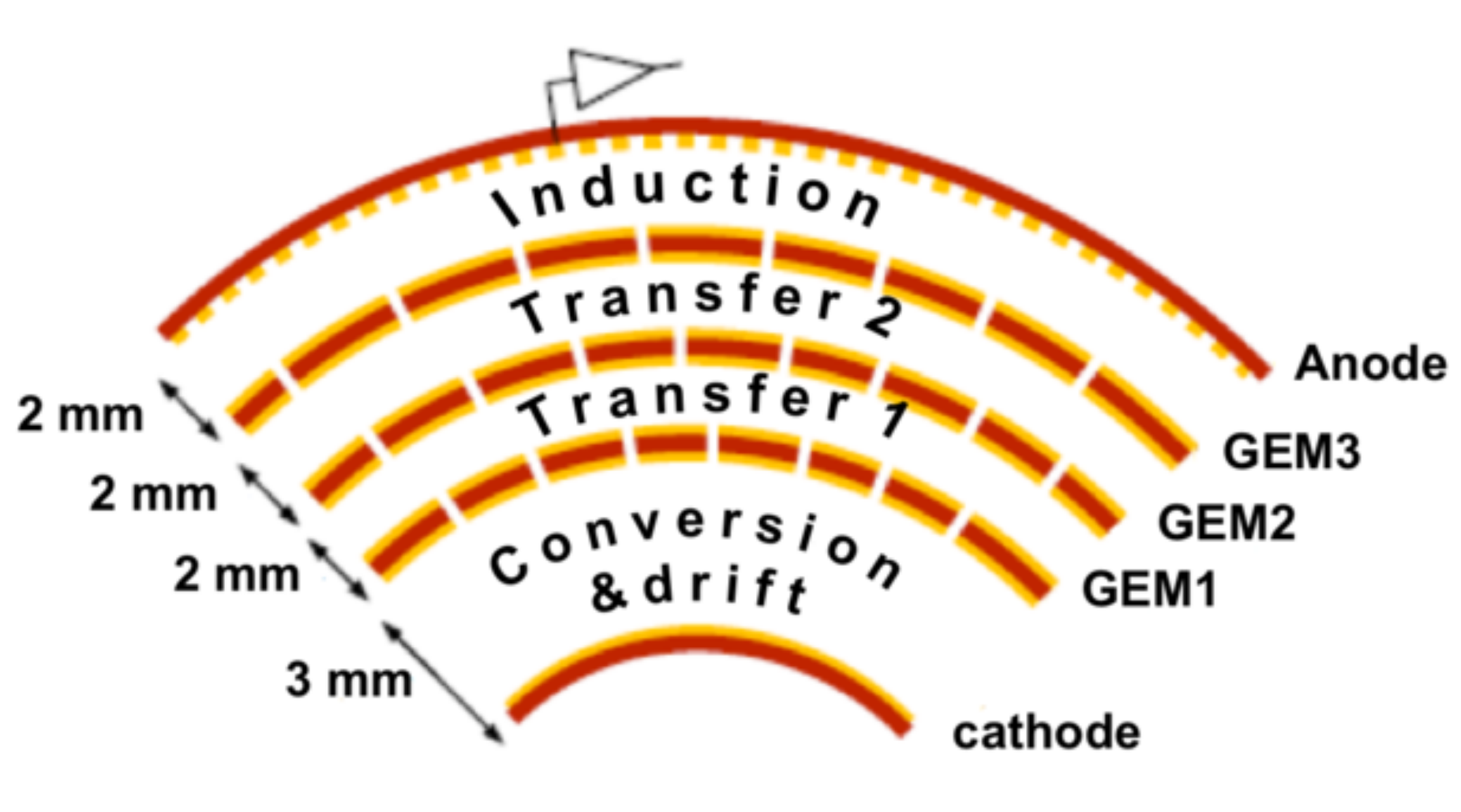}
   		\caption[Schematic representation of the different planes.]{A schematic representation of the different planes in each layer from KLOE-2.}
	\label{fig:triplegem}
	\end{figure}

The KLOE-2 Collaboration built the only existing cylindrical GEM based detector \cite{Balla1,Balla2} consisting of 4 cylindrical tracking layers. Each layer consisting of 5 coaxial cylindrical electrodes, all made of kapton: the cathode, the three GEM stages and the anode readout, see Fig. \ref{fig:triplegem}. Due to GEM foil area limitations the cylindrical electrodes are realized gluing three identical foils, applying epoxy on a 3 mm wide region. The large foil is then wrapped on an aluminium mandrel coated with a very precisely machined 400 $\mu$m thick Teflon film. In order to keep the electrode on the mandrel, during the curing of the epoxy, a vacuum bag system is used to obtain the cylindrical shape given by the mandrel. Anode and cathode electrodes are mounted on a Honeycomb structure that provides the structural support to the detector.

The mechanical support of the chamber is composed by annular flanges made of permaglass placed on the edges of the cylinder. 

The gaps geometry is 3/2/2/2 and the filling gas the Argon-CO$_2$. The detector is placed around the beam-pipe and it is working in a magnetic field of 0.52 T. The operating gain is $2 \times 10^4$ and the readout is formed by $XV$ strips with a 650 $\mu$m pitch and it is equipped with a dedicated chip: GASTONE $\cite{gastone}$, a 64 channels chip featured by a low input equivalent noise and a low power consumption, gave a monostable stretched digital signal. 

The $XV$ stereo angle depends on the layer geometry. It is alternatively positive and negative to reduce the combinatorial background in multi-tracks events. The full system consists of about 10$^4$ electronics channels.\\

The BESIII CGEM IT will take some of the features of these two detectors: it will use an analog readout of the charge deposited by the ionizing particle as COMPASS with a cylindrical shape used in the KLOE-2 CGEM. The main differences with the previous GEM detectors are given by the requirements of the BESIII experiment: an r-$\phi$ spatial resolution at the level of 100 $\mu$m combined with a magnetic field of 1 T.

\section{BESIII innovations}
\label{sec:Innovations}

With respect to the state of art, few important innovations are implemented in the BESIII design.

The innovative aspects are mainly related to the following three items:
\begin{itemize}
\item lighter material for the mechanical structure;
\item the anode design;
\item analog readout mode.
\end{itemize}

\subsection{Rohacell}
The manufacture of the anode and cathode structures adopt a new technique to minimize the material budget with respect to the current state of the art of the detectors. The Rohacell $\cite{rohacell}$ is the trade name of a PMI-based structural foam that is used in fiber composite technology and whose main characteristics are excellent mechanical properties over a wide temperature range, even at low densities, high temperature resistance up to 220 $^{\circ}$C, unique compressive creep behavior for processing up to 180 $^{\circ}$C and 0.7 MPa, excellent dynamic strength and cell sizes that can be tailored for each processing method. Rohacell’s homogenous cell structure provides zero print through to the composite face sheets leaving a class A surface finish every time.

Rohacell is avaible in different densities. During the construction in Ferrara we used 31 IF/IG-F type, which is the lightest one, having a density of 32 kg/m$^3$ and available in foils of different dimensions and thickness. A thickness of 2 mm is suitable for the application of the cathode and 4 mm for the anode.\\

Building the supports for the sub-layers made of Rohacell insted of the Honeycomb allows to reduce the total radiation length significantly while maintaining the mechanical robustness. Moreover the Rohacell is more homogenous than the Honeycomb and that is another advantage for the tracking system.

The combination of Rohacell support structure and the anode configuration makes the CGEM IT an incredibly light detector with a total radiation length of about 1\% of $X_0$. Additional contribution to the $X_0$ comes from a Faraday cage (0.0042 of $X_0$). The material budget calculation is reported in Tab. $\ref{table:X0a}$, $\ref{table:X0b}$ and $\ref{table:X0c}$.

\begin{table}[ht]
\begin{minipage}{\textwidth}
\begin{center}
\begin{tabular}{lccc}
\hline
\textbf{Matherial}	& \textbf{Thickness ($\mu$m)}	& \textbf{Fill factor}	& \textbf{Radiation length (\% of $X_0$)}	\\
\hline
Kapton		& 12.5	& 1 	& 0.004375	\\
\hline
Rohacell 	& 1000	& 1	& 0.007		\\
\hline
Kapton	 	& 12.5	& 1	& 0.004375	\\
\hline
Rohacell 	& 1000	& 1	& 0.007		\\
\hline
Kapton	 	& 12.5	& 1	& 0.004375	\\
\hline
Kapton		& 50	& 1	& 0.0175	\\
\hline
copper		& 3	& 1	& 0.021		\\
\hline
Total 		& /	& /	& 0.065625 	\\
\hline
\end{tabular}
\caption[Material budget calculation for the CGEM-IT cathodes.]{Material budget calculation for the CGEM-IT cathodes.}
\label{table:X0a}
\end{center}
\end{minipage}
\end{table}

\begin{table}[ht]
\begin{minipage}{\textwidth}
\begin{center}
\begin{tabular}{lccc}
\hline
\textbf{Matherial}	& \textbf{Thickness ($\mu$m)}	& \textbf{Fill factor}	& \textbf{Radiation length (\% of $X_0$)}	\\
\hline
copper		& 3	& 0.8 	& 0.0168	\\
\hline
Kapton	 	& 50	& 0.8	& 0.014		\\
\hline
copper		& 3	& 0.8 	& 0.0168	\\
\hline
Total 		& /	& /	& 0.0476 	\\
\hline
\end{tabular}
\caption[Material budget calculation for the CGEM.]{Material budget calculation for the CGEM.}
\label{table:X0b}
\end{center}
\end{minipage}
\end{table}

\begin{table}[ht]
\begin{minipage}{\textwidth}
\begin{center}
\begin{tabular}{lccc}
\hline
\textbf{Matherial}	& \textbf{Thickness ($\mu$m)}	& \textbf{Fill factor}	& \textbf{Radiation length (\% of $X_0$)}	\\
\hline
copper		& 3	& 1 	& 0.021		\\
\hline
Kapton	 	& 50	& 1	& 0.014		\\
\hline
Rohacell 	& 2000	& 1	& 0.014		\\
\hline
Kapton	 	& 12.5	& 1	& 0.004375	\\
\hline
Rohacell 	& 2000	& 1	& 0.014		\\
\hline
Kapton	 	& 50	& 1	& 0.0175	\\
\hline
copper		& 3.5	& 0.87 	& 0.021315	\\
\hline
Kapton	 	& 50	& 0.2	& 0.0035	\\
\hline
copper		& 3.5	& 0.2	& 0.0049	\\
\hline
Total 		& /	& /	& 0.11809 	\\
\hline
\end{tabular}
\caption[Material budget calculation for the CGEM-IT anodes.]{Material budget calculation for the CGEM-IT anodes.}
\label{table:X0c}
\end{center}
\end{minipage}
\end{table}

\subsection{Anode design}
\label{sec:anodesign}
The anode design is developed on the basis of the readout used by the Compass experiment with some improvements.

The readout anode circuit is manufacted from 3 $\mu$m copper clad over 50 $\mu$m thick polymide substrate, the same used for GEM foils. Two foils with copper segmented in strips are used to have the two dimensional readout. The strip pitch is 650 $\mu$m, 570 $\mu$m wide $X$-strips are parallel to the CGEM axis, providing the r-$\phi$ coordinates; while the V-strips, having a stereo angle with respect to the $X$-strips, are 130 $\mu$m wide and, together with the other view, gives the z coordinate. The stereo angle depends on the layer geometry and will be 45.9 $^{\circ}$ for layer 1, -31.1 $^{\circ}$ for layer 2 and 33.0 $^{\circ}$ for layer 3.\\

\begin{figure}[btp]
		\centering
		\begin{subfigure}[h]{0.3\textwidth}
   			\includegraphics[width=\textwidth]{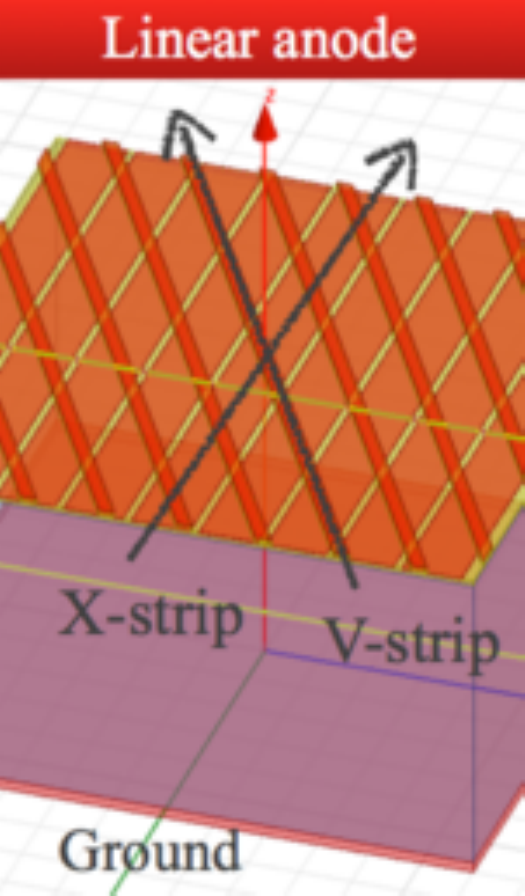}
			\caption{}
			\label{fig:jagged:a}
		\end{subfigure}
		\begin{subfigure}[h]{0.3\textwidth}
   			\includegraphics[width=\textwidth]{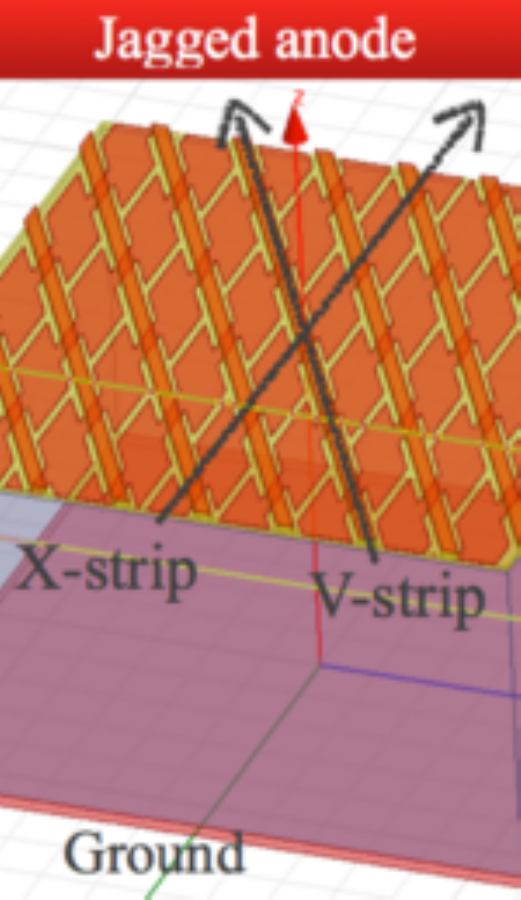}
			\caption{}
			\label{fig:jagged:b}
		\end{subfigure}
   		\caption[Anode layout.]{Anode layout for $XV$ strips with straight (a) and jagged strips (b) that minimizes the capacitance coupling.}
	\label{fig:jagged}
	\end{figure}

\begin{figure}[btp]
		\centering
		\begin{subfigure}[h]{0.4\textwidth}
   			\includegraphics[width=\textwidth]{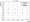}
			\caption{}
			\label{fig:jaggedsim:a}
		\end{subfigure}
		\begin{subfigure}[h]{0.4\textwidth}
   			\includegraphics[width=\textwidth]{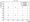}
			\caption{}
			\label{fig:jaggedsim:b}
		\end{subfigure}
   		\caption[Capacity simulations of different anode layout.]{Simulations of the strip capacity for for prototype of different dimensions in linear (blue) and jagged (red) layout between two single adjacent strips (a) and $X$vs$V$ strip (b).}
	\label{fig:jaggedsim}
	\end{figure}

A new layout, with jagged-strip as shown in Fig. $\ref{fig:jagged}$ on the left side, has being studied to minimize the strip capacitance with respect to the linear-strip configuration, shown in Fig. $\ref{fig:jagged}$ on the right side, by about 30\% due to the reduction of the overlapping area between the strips of different views (as shown in Fig $\ref{fig:jaggedsim}$) $\cite{CDR}$. The comparative analysis of the capacitance couplings of the two configurations is performed by means of Maxwell simulations $\cite{maxwell}$.

Another difference respect to the KLOE-2 design is the ground plane distance, increased from 0.2 mm to 2 mm in order to reduce the capacitance coupling between the strip planes and the reference voltage. A Rohacell structure will fill the space between the readout plane and the ground.

\subsection{Analog readout}
\label{sec:anal}
Digital and analog readout have been investigated to reach the required spatial resolution $\le$ 100 $\mu$m.
The latter identifies clusters by detecting adjacent strips with a signal above a fixed threshold; the reconstructed position of the track is the geometrical center of the cluster and the resolution is (roughly) defined by the ratio $pitch/\sqrt{12}$.
The analog readout method, on the other hand, allows to set both a threshold on the single strips and a threshold on the total charge then improving the ghost hit rejection (a ghost hit is an over threshold cluster due to noise). Moreover, the strip collected charge encoding allows the reconstruction of the charge centroid, then boosting the resolution above the $pitch/\sqrt{12}$ ratio of the binary readout method.
The binary readout front-end manages a simple single bit information per strip while the analog readout front-end requires the encoding of the collected charge and a more complex readout chain.

\section{Mechanical design}

The mechanical design that provides the detector support of the three CGEM layers starts by the cylindrical electrodes structure.
To obtain a cylindrical shape a different mandrel is used for each electrode, 15 in total. Detail of the five electrodes and the assembly technique are described in the following sections.

\subsection{Detector elements}

\subsubsection{Cathode}

The cathode is the innermost electrode. The foils are produced by the CERN EST-DEM workshop as polyimide foils, 50 $\mu$m thick, with a copper cladding of 3 $\mu$m on the internal surface. Layer 1 needs one foil and the other layers two. The cathode cylindrical structure made of two 1-mm Rohacell layers. Rohacell foils, cut in helicoidal shape, are rolled around the mandrel; two helicoids, left- and right-handed, are needed to give the robustness of the structure. A 12.5 $\mu$m thin Kapton foil is placed between Rohacell layers to allow a better gluing.

The Rohacell is machined very precisely to 1 mm of thickness through a lathe. The cathode is glued on top of this structure. Annular flanges of permaglass placed on the edges of the cylinder provide the mechanical support of the chamber. These rings house the gas inlets and outlets and their thickness defines the space of the gap between the cathode and the first GEM.\\

\subsubsection{GEM}

In layers 2 and 3, two GEM foils are spliced together in order to realize one single electrode.
In order to limit the capacitance and hence the energy released through the GEM hole in case of discharges, each foil has independent high voltage sectors. Each foil is divided in sectors, 40 for layer 1 and 2, and 60 for layer 3.\\ 

A special assembling technique has been developed to obtain cylindrical GEM electrodes: two GEM foils are glued together on a plane to obtain the single large foil needed to make a cylindrical electrode. Epoxy is applied on one of the sides of the GEM foil, then the foil is rolled on an aluminum mandrel coated with a very precisely machine. The mandrel is inserted in a vacuum bag and the vacuum provide the uniform pressure over the whole surface. The other electrodes (anode and cathode) are built similarly.\\

\subsubsection{Anode}

The anode is manufactured by the CERN EST-DEM as two layers of 50 $\mu$m thick Kapton foils with 3.5 $\mu$m copper strips, providing a 2-D readout. The strip pitch is 650 $\mu$m and is the same for $X$ and $V$ view\footnote{$V$ is the diagonal coordinate. The angle is given by the detector geometry}. The azimuthal coordinate is provided by the $X$-strips with a width of 570 $\mu$m and the other coordinate is given by the $V$-strips, with a width of 130 $\mu$m. The intersection between the $X$ and $V$ strips has a stereo angle that depends on the layer geometry. The strip dimensions are chosen in order to equally share the charge on the strips of the two views, while their pitch allows the achievement of the required spatial resolution keeping the number of channels manageable. The anode plane is rolled on the mandrel and then two 2-mm Rohacell layers are glued on top of it, with a 12.5 $\mu$m Kapton foil between them. At the end the ground plane is glued. As for the cathode and for the GEM, also for the anode a special set of Permaglass annular rings is placed on the edges of the cylinder.

\subsection{Assembly technique}

A special assembling technique has been developed to obtain cylindrical GEM electrodes: two GEM foils are glued together on a plane to obtain the single large foil needed to make a cylindrical electrode. Epoxy is applied on one of the sides of the GEM foil, then the foil is rolled on an aluminum mandrel. The mandrel is inserted in a vacuum bag and the vacuum provide the uniform pressure over the whole surface. The other electrodes (anode and cathode) are built similarly. The five electrodes are extracted from the mandrels along the vertical direction by using a PVC ring, fixed with pins to one of the annular flanges of the cylinder, and then are inserted one into the other. To accomplish the insertion of the electrodes without damaging the GEMs, a dedicated tool has been realized: the Vertical Insertion System (see Fig. \ref{fig:clessidra}).

\section{Cathode's structure construction at Ferrara}
\label{sec:Construction}

\begin{figure}[btp]
		\centering
		\begin{subfigure}[h]{0.4\textwidth}
   			\includegraphics[width=\textwidth]{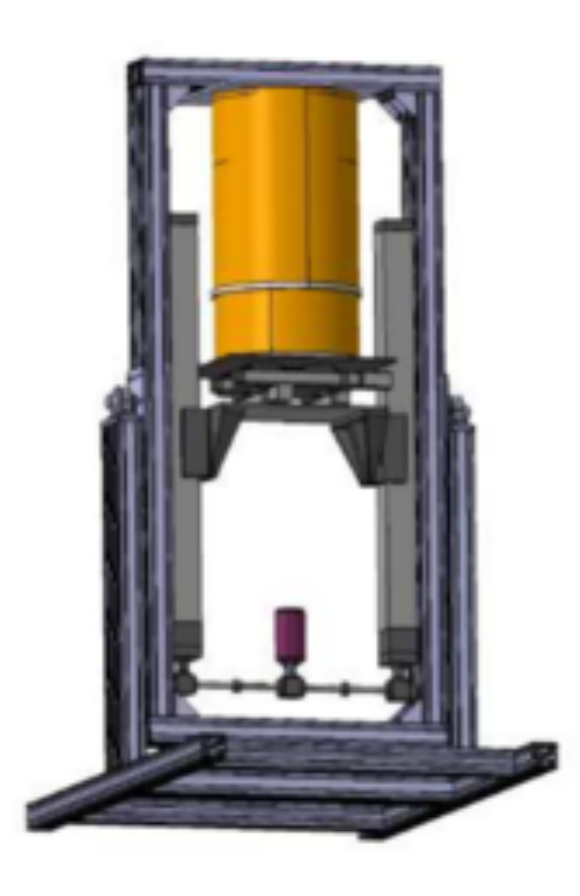}
			\caption{}
			\label{fig:clessidra:a}
		\end{subfigure}
		\begin{subfigure}[h]{0.3\textwidth}
   			\includegraphics[width=\textwidth]{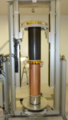}
			\caption{}
			\label{fig:clessidra:b}
		\end{subfigure}

   		\caption[Isometric view of the Vertical Insertion System.]{Isometric view of the Vertical Insertion System, drawing (a) and picture (b).}
	\label{fig:clessidra}
	\end{figure}

The detector manufacture starts from the layer 2. The 5 different sub-layers (chatode, 3 GEMs, anode) will be constructed separately and then it will be assembled together by the specific vertical insertion system, shown in Fig. $\ref{fig:clessidra}$.

\begin{figure}[btp]
		\centering
		\begin{subfigure}[h]{0.22\textwidth}
   			\includegraphics[width=\textwidth]{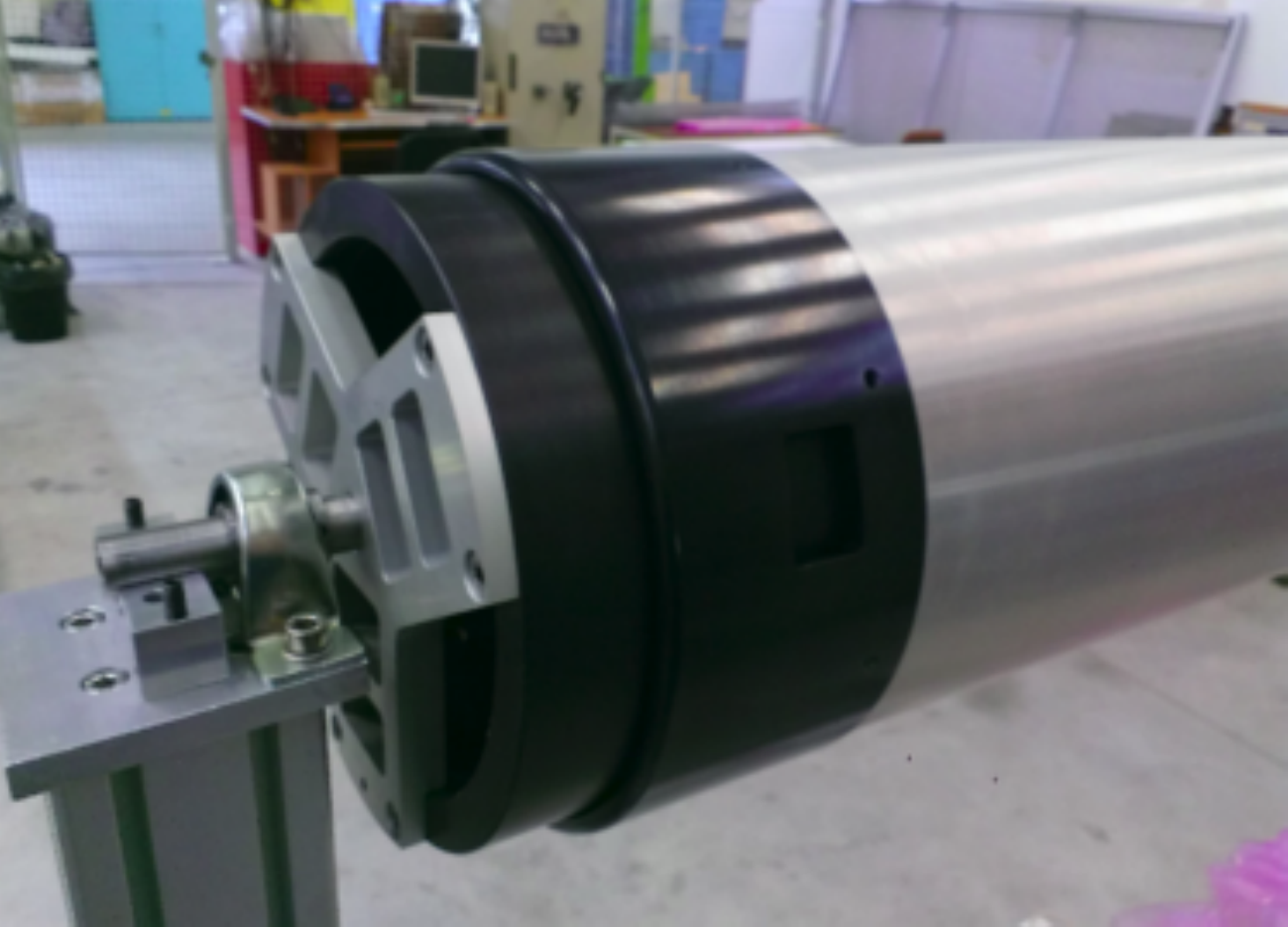}
			\caption{}
			\label{fig:constrlayers:a}
		\end{subfigure}
		\begin{subfigure}[h]{0.6\textwidth}
   			\includegraphics[width=\textwidth]{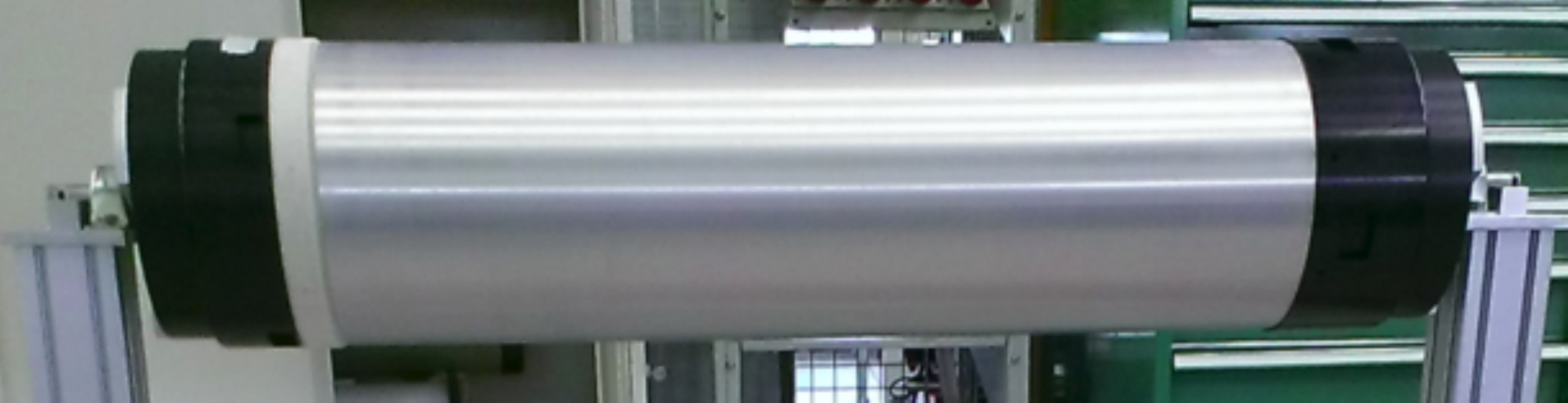}
			\caption{}
			\label{fig:constrlayers:b}
		\end{subfigure}
		\begin{subfigure}[h]{0.68\textwidth}
   			\includegraphics[width=\textwidth]{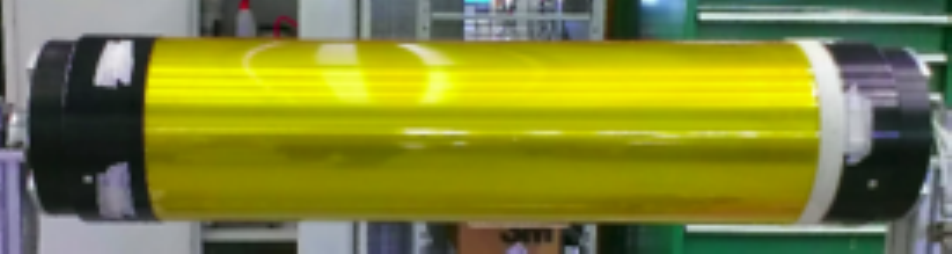}
			\caption{}
			\label{fig:constrlayers:c}
		\end{subfigure}
		\begin{subfigure}[h]{0.68\textwidth}
   			\includegraphics[width=\textwidth]{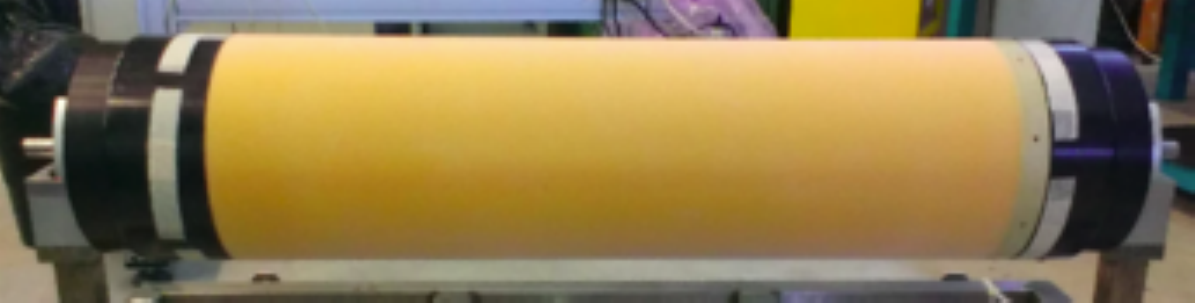}
			\caption{}
			\label{fig:constrlayers:d}
		\end{subfigure}
		\begin{subfigure}[h]{0.68\textwidth}
   			\includegraphics[width=\textwidth]{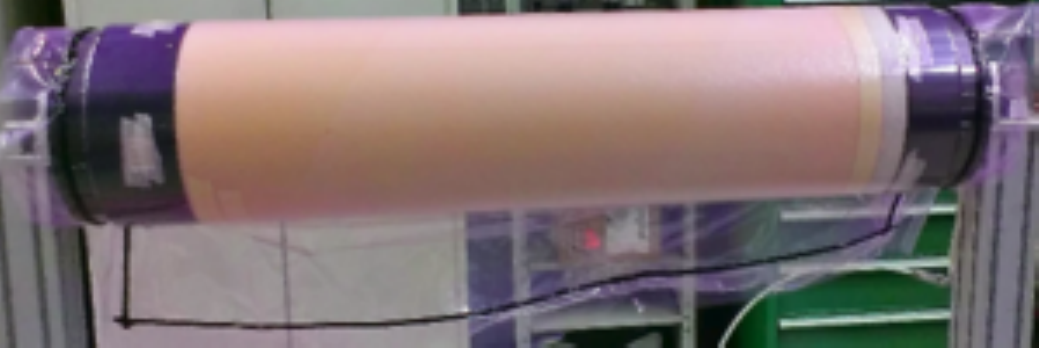}
			\caption{}
			\label{fig:constrlayers:e}
		\end{subfigure}
		\begin{subfigure}[h]{0.68\textwidth}
   			\includegraphics[width=\textwidth]{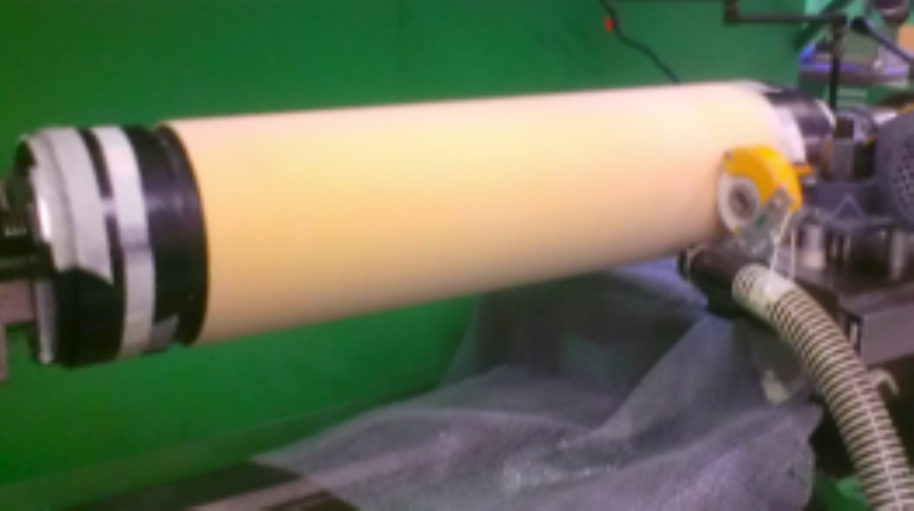}
			\caption{}
			\label{fig:constrlayers:f}
		\end{subfigure}
   		\caption[Pictures of different stages during the construction]{a) and b) the aluminium mandrel with the teflon treatment, the permaglass and the two aluminium rings. c) Kapton foil. d) Rohacell foils. e) The vacuum bag during the gluing procedure. f) The working lethe at INFN-Legnaro.}
	\label{fig:constrlayers}
	\end{figure}

The whole cathode structure of layer 2 is manufactured in Ferrara. The mandrel is the fundamental tool, shown in Figs. $\ref{fig:constrlayers:a}$ and $\ref{fig:constrlayers:b}$: a cylinder of aluminium with a teflon layer that works as a mold for the structure. It consists of a cylindrical surface. 820 mm in length and a 770 mm in circumference, its dimensions begin determined by the inner radius of the layer 2, plus two aluminium rings and an internal gas line used by the vacuum system. An anodizing and teflon treatment is performed to prevent the oxidation.

In the vacuum technique the mandrel is covered with a cylindrical vacuum bag and the air dispersion is sealed up with butyl rubber. The vacuum is pumped from inside down to few mbar, equivalent to $\sim$ 1 kg/cm$^2$, for 12 hours. Using this technique the gluing procedure is very successful and it allows to obtain a very precise cylindrical shape for the Kapton and Rohacell foils with a tolerance of 1 $^{\circ}/_{\circ\circ}$.

	\begin{figure}[btp]
		\centering
   		\includegraphics[width=0.8\textwidth]{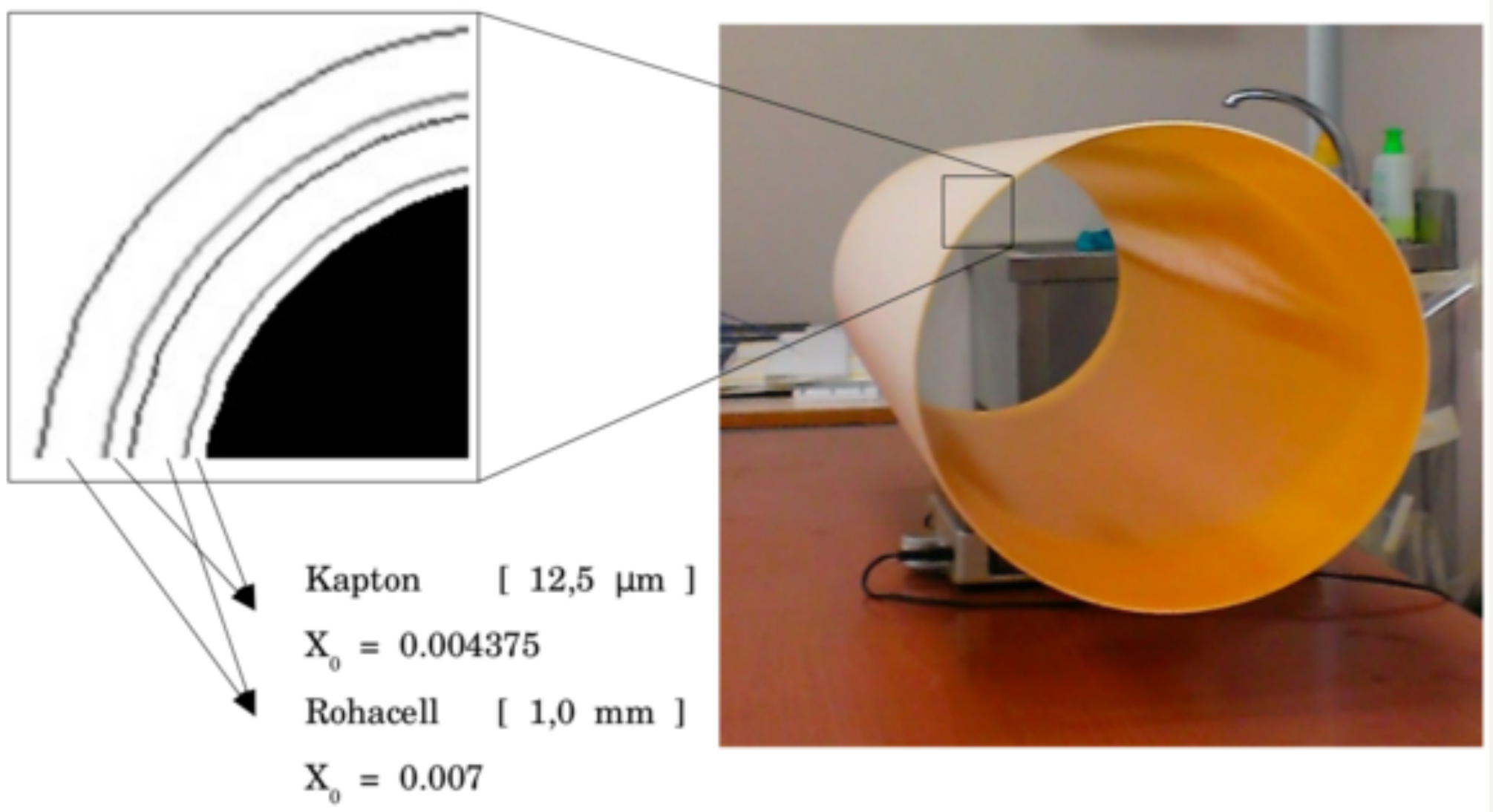}
   		\caption[Cathode stratigraphy.]{Stratigraphy of the cathode of the layer 2 built in Ferrara (right) with a drawing of the different layers used (left) and their dimensions and radiation leghts.}
	\label{fig:stratigraphy}
	\end{figure}

The Rohacell foils need to be prepared before the use: any dust is removed by means of a hoover then they are cut from 2500 $\times$ 1200 mm$^2$ surface to a helicoidal shape.
The two Rohacell foils are rolled around the molds, once clockwise and then anti-clockwise in order to have a higher rigidity.
For a better gluing, a 12.5 $\mu$m thin, Kapton foil is placed between the Rohacell and the other surfaces (mandrel, another Rohacell foil, electrode). 

The stratigraphy from the inner side to the outer is Kapton, Rohacell, Kapton, Rohacell and electrode. 

The gluing is performed by a transfer technique with Mayer foil with an Araldite-based epoxy. During the gluing, each layer is fixed around the mandrel with the vacuum pump and the cylindrical bag. 

The standard Rohacell foil thickness is 2 mm. To reduce it to 1 mm a specific lathe is used, with a precision of tens $\mu$m, at INFN laboratory in Legnaro.

The final result, without the electrode, weighs 180 g and it is shown in Fig. $\ref{fig:stratigraphy}$.Fig. $\ref{fig:constrlayers}$ shows the different steps of the construction: on the top the mandrel, the first Kapton layer, the Rohacell layer, the vacuum technique and on the bottom a picture during the operation with the lathe.

\section{Front-end electronics}

The front-end electronics will be located on the detector, then both dimensions and power dissipation are issues that must be considered in the design. The board dimensions are mainly dominated by the strip connector and the preamplifier (ASIC) protection circuit, while the power dissipation is strictly related to the ASIC circuit. Therefore, the single channel power consumption must be as lower as possible .

Each preamplifier boards will host two ASICs, 64 channels each one, for strips readout. The devices must provide a low sensitivity to the input capacitance ($i.e.$ a series noise contribution of the order of 40 erms/pF), as well as good gain (of the order of 4-5 mV/fC in the head-stage preamplifier) and the capability to sustain the foreseen single channel rate.

Although, the wide spread of the input capacitance (due to the different $V$ strips lengths) does not allows a fully optimization of the head/shaping stages, efforts have been devoted to minimize the readout electrode parasitics and, as a consequence, the parasitic spread. Because the strip capacitances are dominated by the two view cross-capacitance, a study has been carried out using Maxwell simulator to investigate the possibility of decreasing the coupling as shown in Sec. \ref{sec:anodesign}.

A further critical point in the front-end chain design consist in the single strip input rate (signal + background). A high input rate would require a baseline restorer before the ASIC analog to digital conversion stage to limit the offset fluctuation and also deeper buffers in the readout boards then increasing the time required to find the event correct window. An estimation of the single strip maximum rate will be presented in Sec. \ref{chBackground}.

The analog readout chosen for the BESIII CGEM-IT requires the development of a custom ASIC, in fact the GASTONE ASIC \cite{gastone} used to instrument the KLOE-2 CGEM-IT cannot be used because delivers only a single bit information per strip.
The moderate charge resolution required by the GEM detector makes the use of the Time-over-Threshold (ToT) technique, already used in \cite{Gmazza,MDrolo}, an attractive solution for data digitization.
A ToT system is very similar to a simple binary one and it consists of a front-end amplifier followed by a threshold comparator. The charge information is preserved by encoding the duration of the discriminator output pulse. When the signal crosses the threshold, the comparator fires and the corresponding time is captured by the digital logic and provides the event time-stamp. The trailing edge of the signal is also stored to allow to measure by difference the total pulse length.
By properly designing the analog circuitry, it is possible to maintain an adequate linearity.

	\chapter{Background studies}
\label{chBackground}

Background studies allow to extract the occupancy of the strips of the CGEM IT which is an important parameter for the detector optimization and for design of the front-end electronics.
Background rates from this study will be used also to perform simulations of real data analysis to evaluate the expected performance of the CGEM-IT.

Experimental investigations show that the beam related background of the MDC in BESIII experiment come from the beam gas interaction and the Touschek effect, which depend on the machine status, especially at high beam currents $\cite{CDR}$.

Real background events are acquired during the data-taking at different energy scale and time. These data are acquired using the random trigger and usually are used to simulate the background during the simulations. Each data run has its own raw random trigger data.

The background level expected for the CGEM detector can be extracted from measurement of the occupancy of each wire in the MDC of the random trigger events scaled by a factor, that depends on the different geometry and material, that can be extracted from a Monte Carlo simulation. 

\section{BESIII Offline Software}
\label{sec:boss}
The BES-III Offline Software System (BOSS) uses the C++ language and object-oriented techniques and runs primarily on the Scientific Linux CERN (SLC) operating system. BOSS provides simulation, reconstruction, calibration and analysis tools. The BOSS framework is based on the Gaudi package \cite{Gaudi}. In the background studies BOSS is used to analyze the data from the random trigger to study the noise level for each MDC layer.

The Bhlumi $\cite{Bhlumi}$ and Bhwide $\cite{Bhwide}$ QED algorithm generators are used to generate the Bhabha scattering process $e^+e^- \rightarrow e^+e^- + n\gamma$. These are full energy scale generators, the Bhlumi generator is suitable for generating low angle Bhabha events ($\theta$ < 6$^{\circ}$), while the Bhwide generator is intended for wide angle Bhabha events ($\theta$ > 6$^{\circ}$). The use of this process is enough for our purpose since the MC is only used to scale the rates from MDC data to the different detector design. The effect of different angular distributions on the scale factor has been checked using different processes for which we found deviations less than 20\%.

A GEANT4 $\cite{geant}$ model of the CGEM IT was already developed with geometry and material close to the final design and a proper BOSS software is being developed from Chinese and Italian team. These tools allow to simulate the interaction between the particles and the CGEM detector.  

\section{Present and expected backgrounds}

Raw data of different runs have been taken into account. To extract the occupancy of the MDC channels, the information about the MDC geometry has been used: 43 layers with a number of wires that goes from 80 (inner layer) to 576 (outer layer), a wire length from 780 mm to 2308 mm and a radial position from 78.85 mm to 763.45 mm.

\begin{figure}[btp]
		\centering
   		\includegraphics[width=0.8\textwidth]{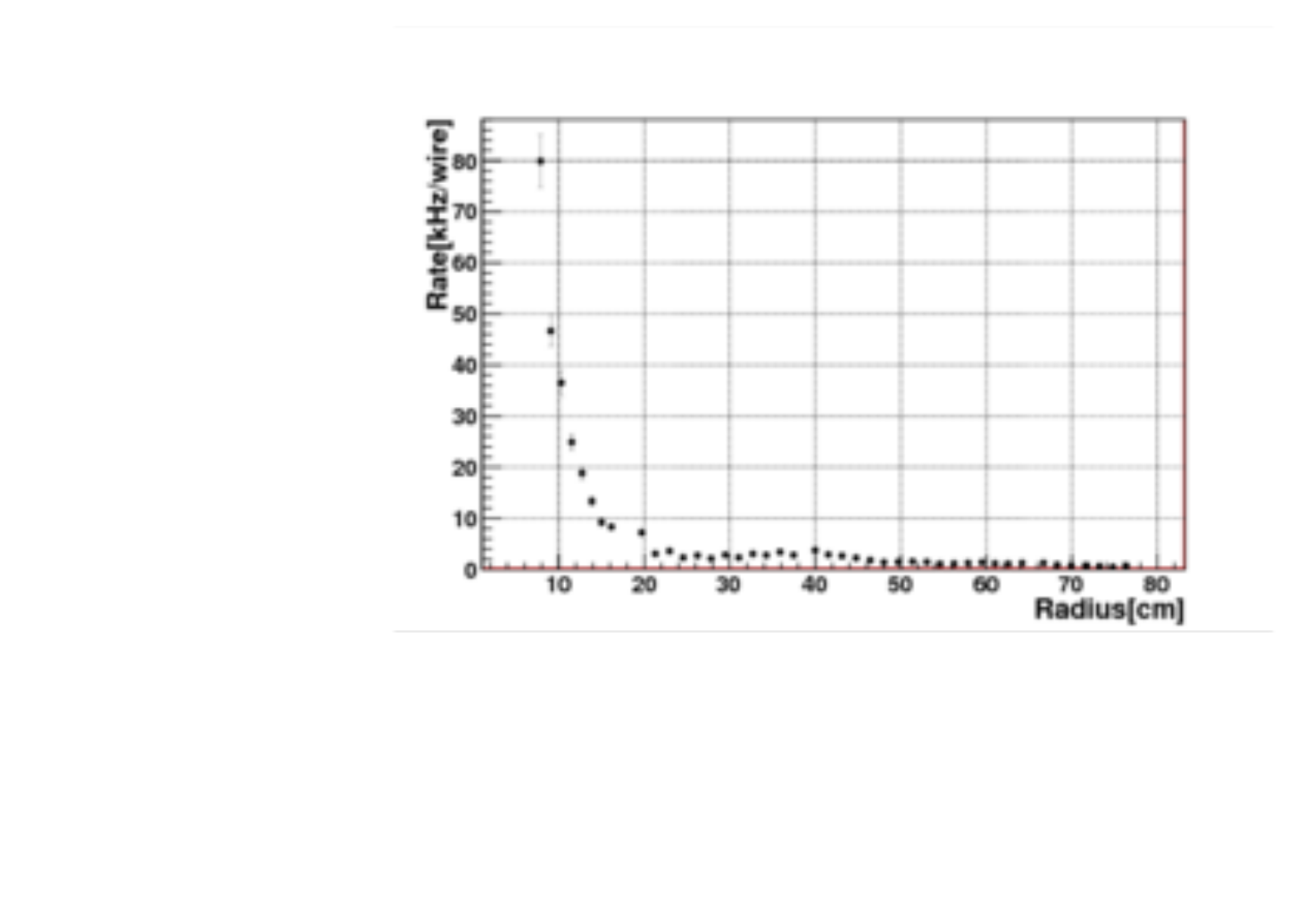}
   		\caption[Rate per wire on the inner drift chamber.]{Average rate per wire on the inner drift chamber as function of the radius. Each point represents a layer. The inner drift chamber is composed by the first eight layers.}
	\label{fig:MDCrate}
	\end{figure}

The trigger window is 2.0 $\mu$s so using the occupancy information from each wire it is possible to calculate the rate/wire for each layer versus the distance from the beam-line (radius). A single raw data is shown in Fig. $\ref{fig:MDCrate}$: the nearest layers are more noisy with respect to the others. The rate/wire on the first layer is around 80 kHz and it corresponds to a density rate on this layer of 0.17 kHz/cm$^2$.

\begin{figure}[btp]
		\centering
   		\includegraphics[width=0.8\textwidth]{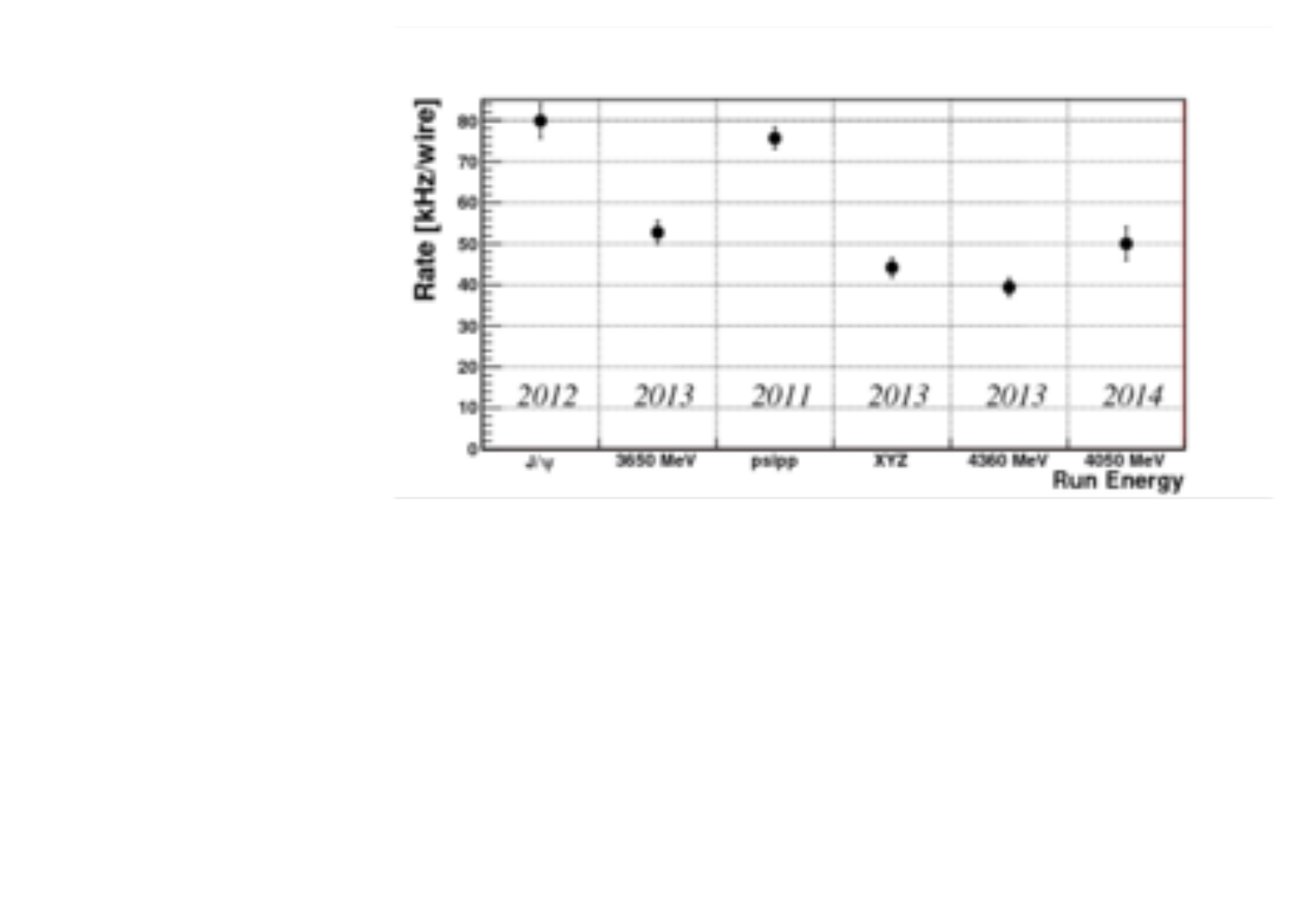}
   		\caption[Rate per wire on the first layer of the drift chamber.]{Average rate per wire on the first layer of the drift chamber for different running periods and energy beams.}
	\label{fig:MDCage}
	\end{figure}

Since the background level is mainly related to machine conditions and beam parameters we calculated the average wire rate of the MDC innermost layer for runs belonging to different recent running periods and energy beams (Fig. $\ref{fig:MDCage}$). The measured rate/wire on the first layer goes from 40 kHz to 80 kHz from 2011 to 2014. Older data are more noisy because of bad condition of the machine background.\\

The expected background rate on the CGEM IT can be estimated by combining the MDC background data with Monte Carlo (MC) simulations. The geometry and the materials difference are taken into account by GEANT4 full simulations. Two sample of MC data performed by the generator described in Sec. $\ref{sec:boss}$ simulate Bhabha events and the two detectors response. Bhlumi and Bhwide allow to check the different effect of different angular distributions: this difference is less than 20\%.

The expected strip rate for the CGEM IT is computed as:

	\begin{equation}
	\label{eq:CGEMrate}
		R^{exp}_{GEM}(r) =  R^{data}_{MDC}(r) \times \frac{R^{MC}_{GEM}(r)}{R^{MC}_{MDC}(r)}
	\end{equation}

where $R^{exp}_{GEM}(r)$ is the expected rate per strip on the CGEM IT as function of the radius. $ R^{data}_{MDC}(r)$ is the MDC wire rate measured from random trigger data and $R^{MC}_{MDC}(r)$ and $R^{MC}_{GEM}(r)$ are the related quantities extracted from the MC simulation. Quantities from the MDC are interpolated to match the radii of the CGEM planes. For the CGEM only the X strips (parallel to the beam axis) are considered. 

\begin{figure}[btp]
		\centering
   		\includegraphics[width=0.8\textwidth]{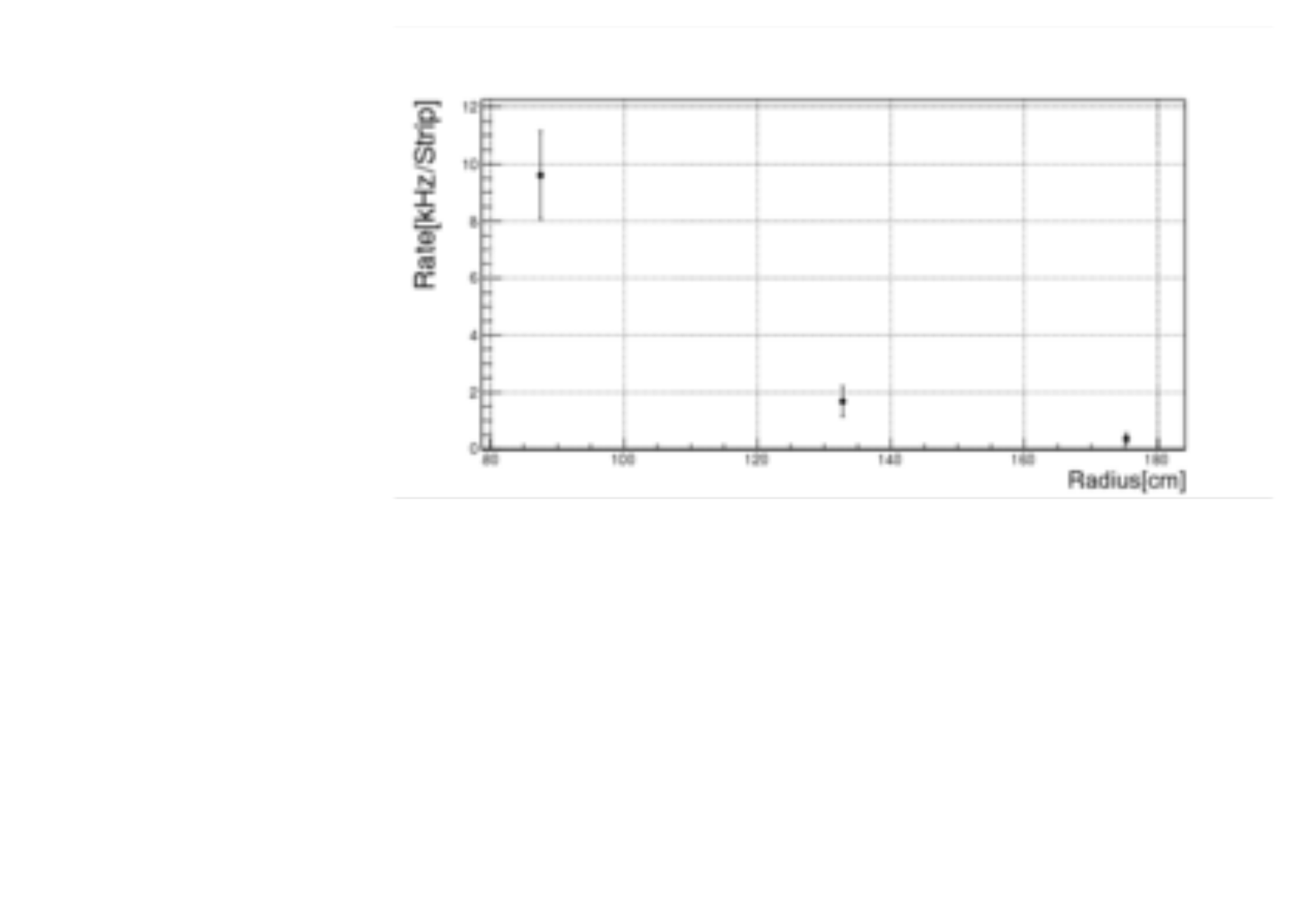}
   		\caption[Expected rate per $X$-strip on the CGEM IT.]{Expected average rate per $X$-strip on the CGEM IT as function the radius.}
	\label{fig:CGEMrate}
	\end{figure}

The values for the three CGEM layers are reported in Fig. $\ref{fig:CGEMrate}$: for the innermost layer the rate/strip is expected to be $\sim$ 9.5 kHz. This result is consistent by scaling the MDC rate/wire by the ratio of the strip/cell size and considering a CGEM strip multiplicity of 3.

A safety factor x6 has to be considered in order to take into account the approximation in the description of the detector in the digitization, systematic error due the distribution and luminosity increase. In conclusion the maximum rate expected for the CGEM IT is about 60 kHz per strip.

	\chapter{CGEM Monte Carlo simulations}
\label{chGarf}

The Monte Carlo simulation is of a paramount importance to understand and to study the properties and the performance of the detector before starting the actual construction. BESIII has a full GEANT4 \cite{geant} simulation describing the detail of the components of the detector and the accelerator elements, and the interaction between the particles and the detector material. The GEANT4 simulation requires input for the hit digitization that has to be provided by the prototype test and more specific simulations of the electron amplification inside the gas volume. The latter is provided by Garfield++, a simulation tool that will be described in Sec. \ref{sec:gargar}.

\section{Garfield modeling of the detector}
\label{sec:gargar}
Garfield++ $\cite{garfield}$ is a toolkit for the detailed computational simulation of detectors which use gases or semi-conductors as sensitive medium. The main area of application is currently in micropattern gaseous detectors like Micromegas and GEM. Garfield++ shares functionality with Garfield, previous version written in FORTRAN, for field simulation and for transport and ionization in gas mixtures. The main differences are the updated treatment of electron transport in gases and the user interface, which is derived from ROOT.

Garfield++ uses several additional tools during the simulation such as:
\begin{itemize}
\item $Magboltz$ that  solves the Boltzmann transport equations for electrons in gas mixtures under the influence of electric and magnetic fields.
\item $Heed$ that generates ionization patterns of fast charged particles. Its core is a photo-absorption and ionization model. Heed in addition provides atomic relaxation processes and dissipation of high-energy electrons.
\item an interface with the finite element program $Ansys$ as basis for Garfield++ electric field calculations.
\end{itemize}

Ansys $\cite{ansys}$ is a finite element simulation software for static, dynamic and thermal problems. It can compute the electric field in the space node-by-node, given a certain geometry and potential at the electrodes. Garfield++ can use as input files the field maps produced by Ansys. Typical node number is $\sim$ 10$^5$, using a mesh-refinement process it is possible to increase the precision.\\

Unfortunately Garfield++ cannot reproduce correctly the gas gain due to the charging up effect that is not properly reproduced by the simulation. Therefore we use it mainly to extract the size and the shape of the electron cloud at the anode to compare different configurations.\\

Due to huge time consumption of such simulations ($e.g.$ a single events where a charged particle interacts with the detector need about a day of computation with a normal computer) a computer farm of 200 cores has been provided by INFN-Torino to run the production.\\
 
In the further sections the simulation results of the gas behaviour and the response of the detector to the charged particles will be shown. The charge distribution of the avalanche electrons at the anode allows to extract the cluster multiplicity and the drift properties of collected charge that will be used as input of the hit digitization.

\section{Gas properties calculation}
\label{sec:gas}

In the CGEM IT the gas is the main object which the charged particles interact: the ionized charge is produced and the coordinate is determined by the measurement of the center of gravity of the collected charge.
The drift velocity and the gas gain differ from gas to gas.

Once the ionization takes place, under the influence of the applied field, electron and ion drift in opposite directions and diffuse towards the electrodes. The scattering cross section is determined by the details of atomic and molecular structure. 

The drift velocity and diffusion of electrons depend very strongly on the nature of the gas, specifically on the inelastic cross-section involving the rotational and vibrational levels of molecules. In noble gases the inelastic cross section is zero  below the excitation and ionization threshold $\cite{ChinisePhysics}$ and the multiplication occurs at much lower fields than in complex molecules. Therefore, convenience of operation suggests the use of a noble gas as the main component. The excited noble gases can return to the ground state only through a radiative process and the minimum energy of the emitted photon (11.6 eV) is above the ionization potential of the metal constituting the cathode (7.7 eV). Then these photons have to be absorbed because new electrons can be extracted and start a new avalanche.\\

Adding polyatomic gases, such as CO$_2$ or C$_4$H$_{10}$, the drift velocities increase because of their large inelastic cross sections at moderate energies, which result in "cooling" electrons into the energy range of the Ramsauer-Townsend minimum ( $\sim$ 0.5 eV ) of the elastic cross-section for Argon. The reduction in both the electron energy and the total electron scattering cross-section results in a large increase of electron drift velocity.

Another role of the polyatomic gas is to absorb the ultraviolet photons emitted by the excited noble gas atoms.\\

Thanks to collections of experimental data $\cite{Peisert}$ and theoretical calculations based on transport theory $\cite{Biagi}$, drift and diffusion properties can be estimates in pure gases and mixtures. As first approximation, the drift velocity $v$ is described as a function of the mean collision time $\tau$ and the electric field E: $v = eE\tau/m_e$ (Townsend's expression). Another theory, the friction force model, provides an expression for the drift velocity $v$ as function of electric and magnetic field vectors \textbf{E} and \textbf{B}, of the Larmor frequency $\omega$ = eB/m$_e$ and of mean collision time $\tau$:

	\begin{equation}
	\label{eq:driftvelocity}
		\textbf{v} = \frac{e}{m_e} \frac{\tau}{1+\omega^2\tau^2} \left( \textbf{E} + \frac{\omega\tau}{B} ( \textbf{E} \times \textbf{B} ) + \frac{\omega^2\tau^2}{B^2} (\textbf{E} \cdot \textbf{B})\textbf{B} \right)
	\end{equation}

The electron velocities and diffusion parameters can be computed with MAGBOLTZ. 

	\begin{figure}[btp]
		\centering
		\begin{subfigure}[h]{0.4\textwidth}
   			\includegraphics[width=\textwidth]{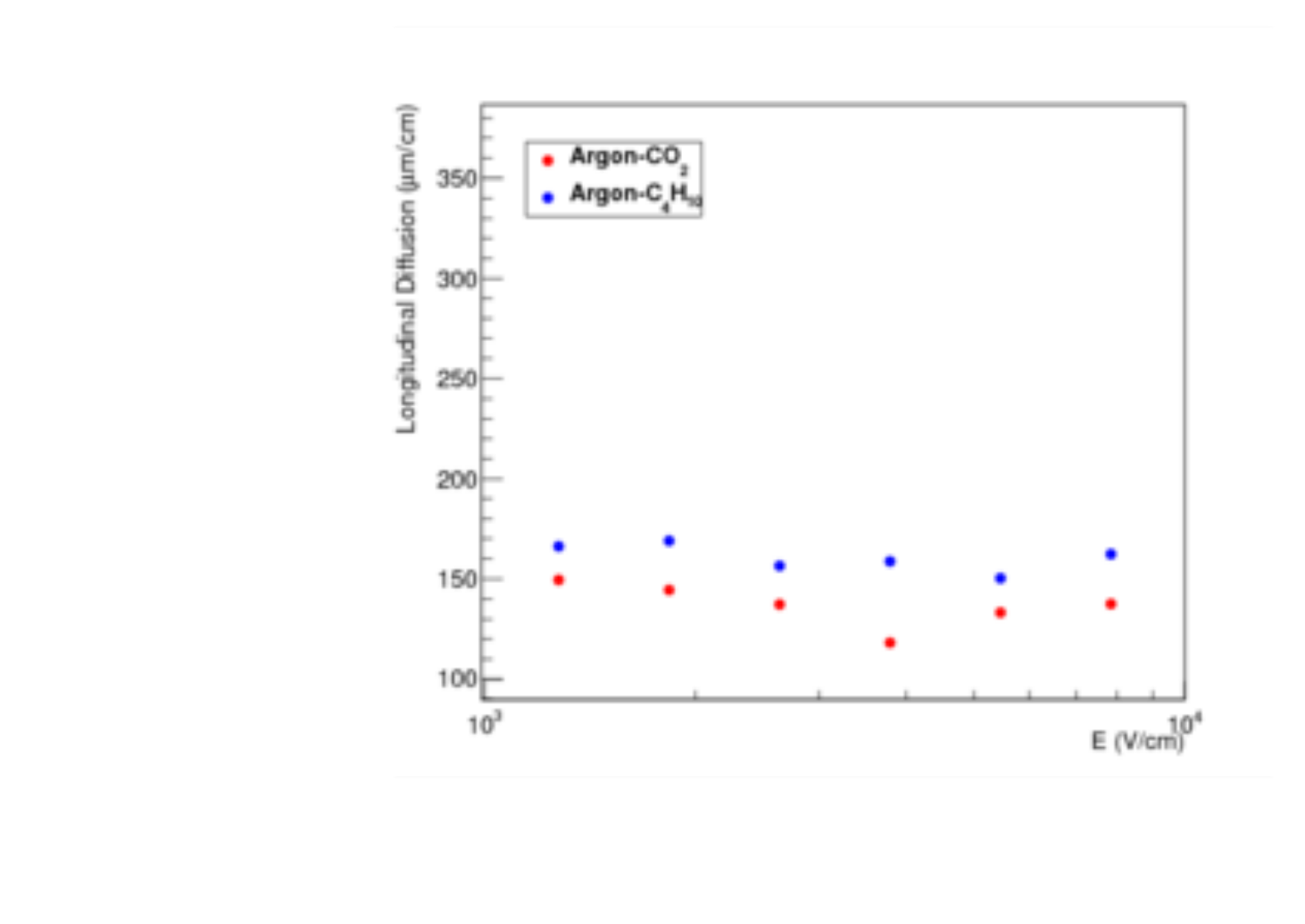}
			\caption{}
			\label{fig:magboltz:a}
		\end{subfigure}
		\begin{subfigure}[h]{0.4\textwidth}
   			\includegraphics[width=\textwidth]{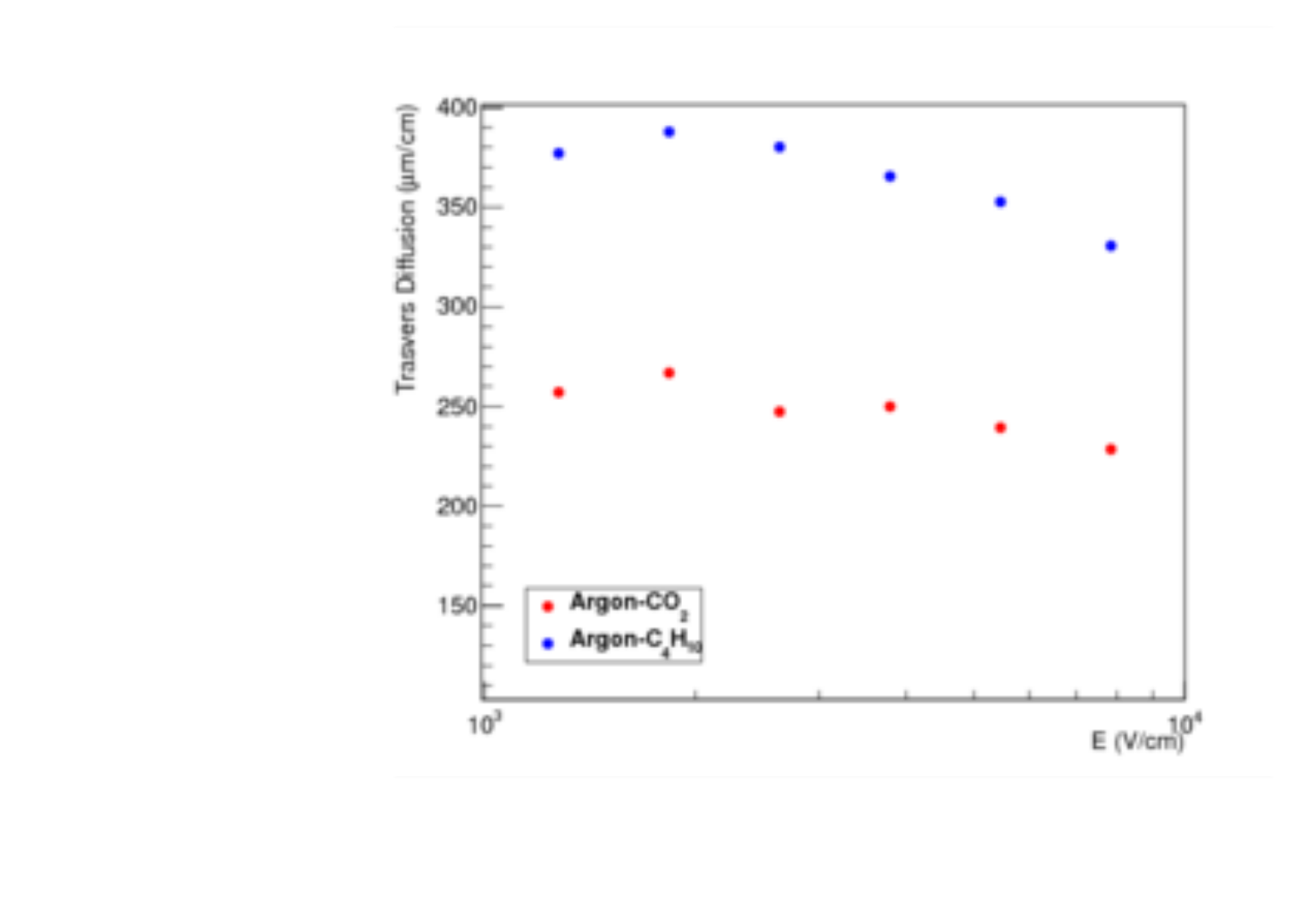}
			\caption{}
			\label{fig:magboltz:b}
		\end{subfigure}
		\begin{subfigure}[h]{0.4\textwidth}
   			\includegraphics[width=\textwidth]{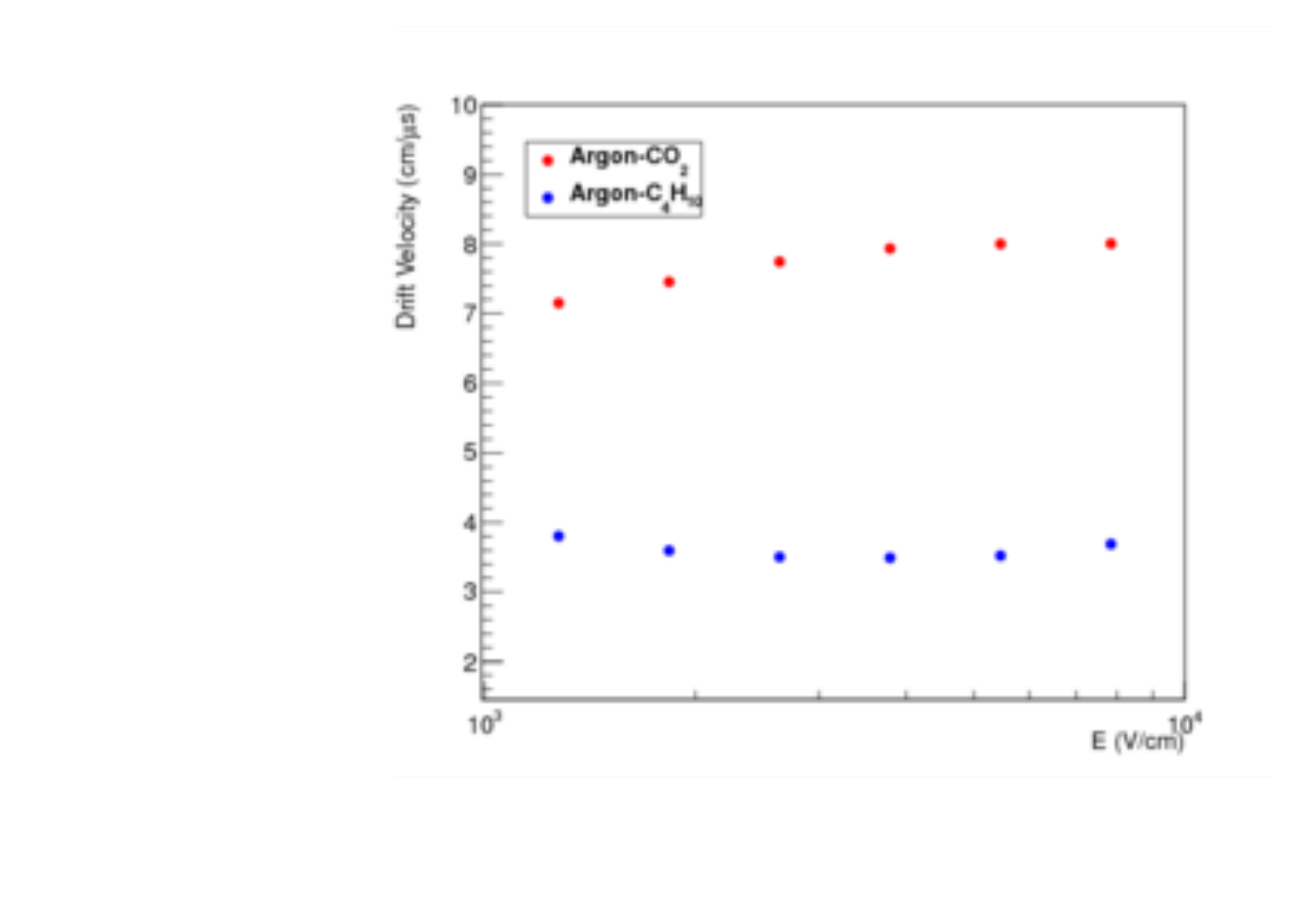}
			\caption{}
			\label{fig:magboltz:c}
		\end{subfigure}
   		\caption[Drift results with Magboltz simulation]{Magboltz simulation results for Argon-CO$_2$ and Argon-C$_4$H$_{10}$ mixtures as function of the electric field. a) Longitudinal drift. b) Transversal diffusion. c) Drift velocity.}
	\label{fig:magboltz}
	\end{figure}

Fig. $\ref{fig:magboltz}$ shows the behaviour of the longitudinal and transverse diffusion, and of the drift velocity as function of the electric field for two different gas mixtures (Argon-CO$_2$ and Argon-C$_4$H$_{10}$). The latter gas mixture is slower than the former and the transverse diffusion is bigger, therefore a larger avalanche size is expected for Argon-C$_4$H$_{10}$ (90:10) \footnote{from now on, Argon-C$_4$H$_{10}$ (90:10) will be referred as Argon-C$_4$H$_{10}$} mixture.

The detected charge depends on the potential differences between the electrodes \cite{sasauli}: 
\begin{itemize}
\item at low voltage, charges begin to be collected but recombination is still the dominant process; 
\item at a certain voltage, called threshold voltage, the field is large and the process of multiplication occurs in a proportionality regime; 
\item at high voltages this proportionality is gradually lost;
\item at even higher voltages starts a region of saturated gain, Geiger-Müller region, which introduces prohibitive dead-times for the electronics.
\end{itemize}

\begin{figure}[btp]
		\centering
		\begin{subfigure}[h]{0.4\textwidth}
   			\includegraphics[width=\textwidth]{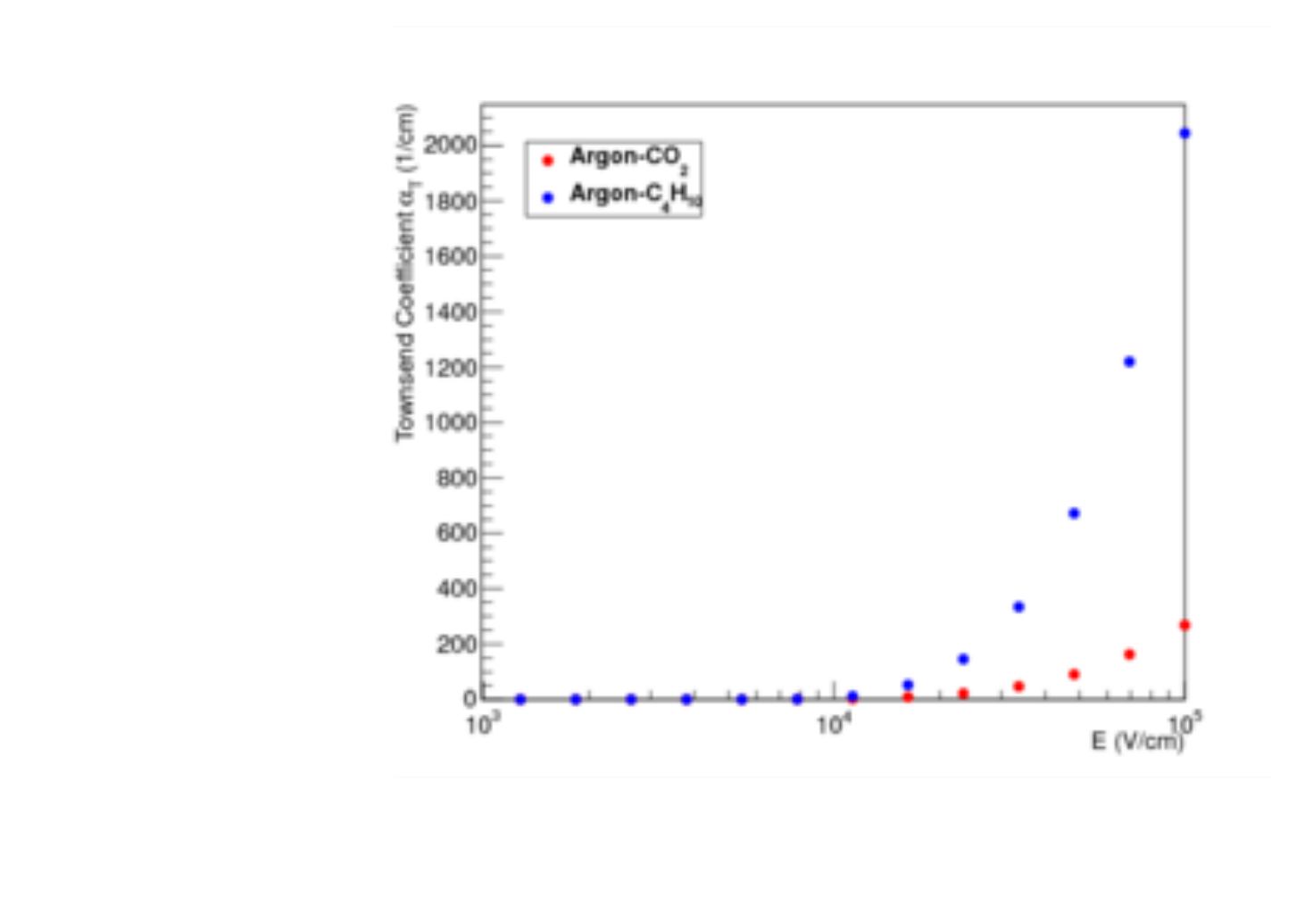}
			\caption{}
			\label{fig:attachtown:a}
		\end{subfigure}
		\begin{subfigure}[h]{0.4\textwidth}
   			\includegraphics[width=\textwidth]{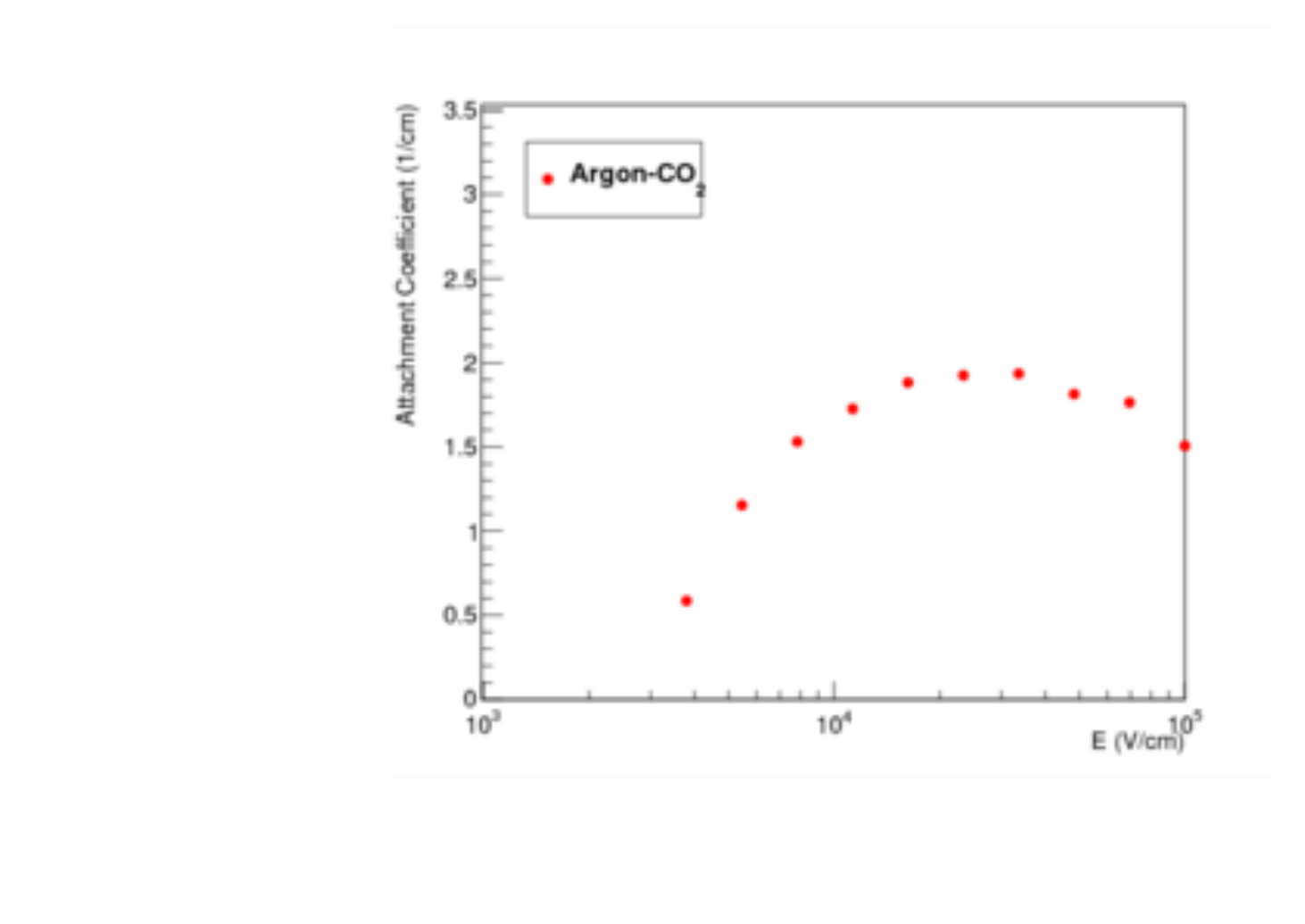}
			\caption{}
			\label{fig:attachtown:b}
		\end{subfigure}
   		\caption[Multiplication coefficient results with Magboltz simulation]{Townsend (a) and Attachment (b) coefficients from Magboltz simulation that describe the main parameters of the electron multiplication in gases.}
	\label{fig:attachtown}
	\end{figure}

Above a gas-dependent threshold, the mean free path for ionization, $\lambda_i$ decreases exponentially with the field; its inverse, $\alpha_T = 1 / \lambda_i$, is the first Townsend coefficient. Townsend coefficient is defined as the number of ions produced per unit path by a single electron. In uniform fields, the electron avalanche size $N$ can be described as a function of $\alpha_T$, the length $x$ and the initial electron number $N_0$: $N = N_0 e^{\alpha_{T} x}$; N/N$_0$ is the gain of the detector. Simulation results are shown in Fig. $\ref{fig:attachtown:a}$\\

In order to work at the highest possible gains, without entering into the Geiger-Müller region, the Argon has to be mixed to other gases because it doesn't allow gains in excess of 10$^3$-10$^4$ \cite{sasauli}. Polyatomic molecules, blocking secondary emission due to Argon recombination with the cathode and absorbing photons, allow to reach gains greater than 10$^4$ without discharge.

During the drift there is a probability that an electron is absorbed by the gas. This effect is described by the attachment coefficient, it corresponds to the probability that an electron drifting through a gas under the influence of a uniform electric field will undergo electron attachment in a unit distance of drift. Simulation results in Fig. $\ref{fig:attachtown:b}$ show that this effect takes place with an electric field between 6 $\times$ 10$^3$ V/cm and 10$^6$ V/cm. No significant dependence on magnetic field is shown.

\section{Triplegem simulations with Garfield++}

\subsection{Ansys cell}

\begin{figure}[btp]
		\centering
   		\includegraphics[width=0.8\textwidth]{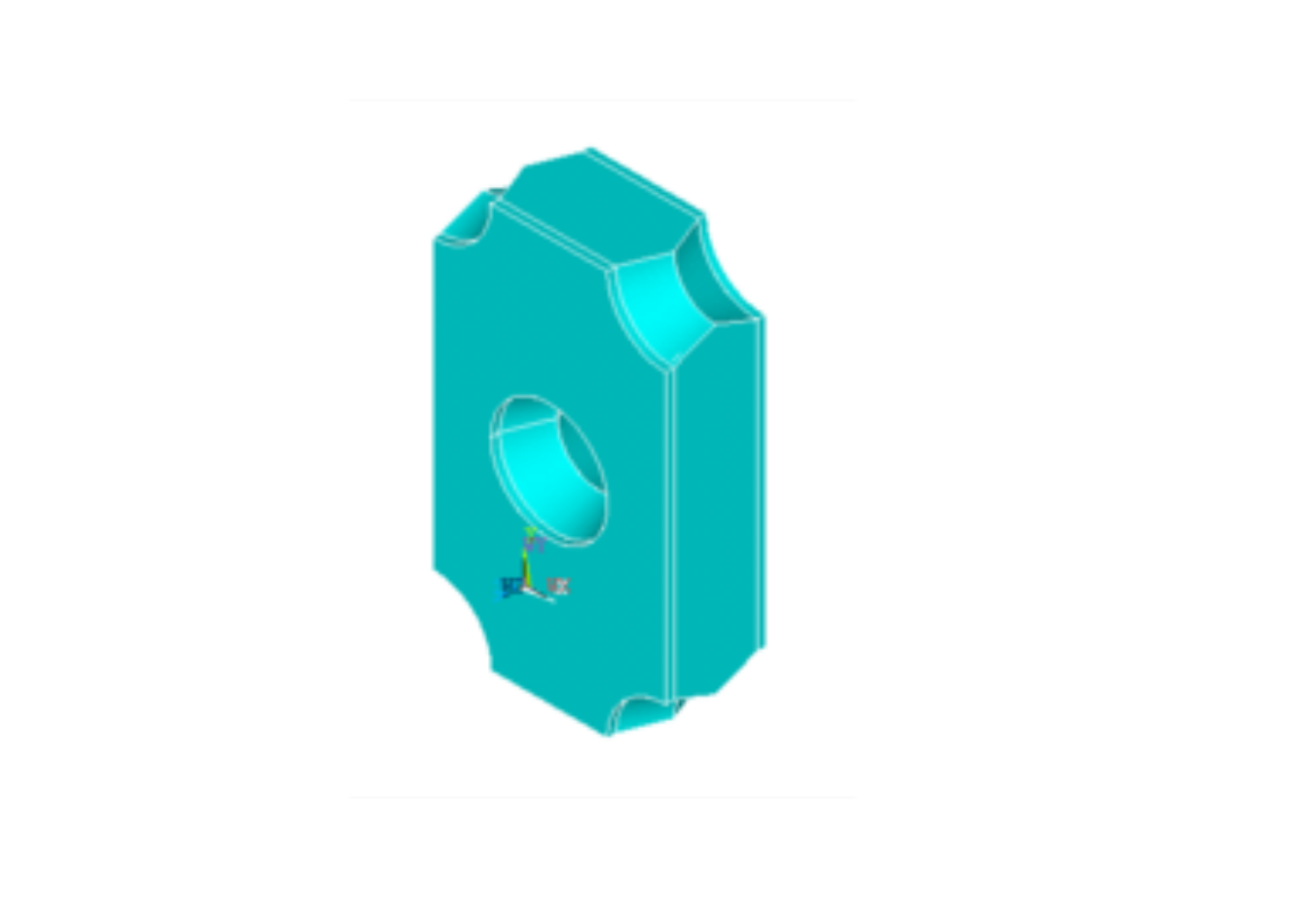}
   		\caption[Ansys GEM hole.]{Ansys GEM cell to replicate along the $x$ and $y$ coordinate to obtain a GEM plane. The geometry recreate iso-separeted holes with biconical shape.}
	\label{fig:ansys}
	\end{figure}

To simulate the CGEM geometry a planar standard of triplegem is considered because the size of involved holes in a single avalanche is very small with respect to the curvature of the cylinder, then the approximation is good. Each hole is biconical with an inner (outer) hole of 55 $\mu$m (70 $\mu$m) and a pitch of 140 $\mu$m. Three close holes create an equilateral triangle so the cell shape is given by Fig. $\ref{fig:ansys}$. This geometry is optimized to maximize the gain and the transparency by data from experiments \cite{Bach2} and simulations \cite{sven}.

Each GEM is composed by a 50 $\mu$m Kapton thickness and 3 $\mu$m copper foils for each side. Three cells are placed at 2 mm distance plus 2 more electrodes for the cathode and the anode. The full geometry is 3/2/2/2 (cathode/ transfer1/ transfer2/ transfer3/ anode). In the simulations, a replication of the cell along $x$ and $y$ is used to create a larger plane. The electric field and the drift direction are along the $z$ coordinate.

\subsection{Fields}

In Sec. $\ref{sec:gas}$ the gas properties as function of the electric field have been studied. These results are used to set the correct field between the electrodes. A field of 1.5 kV/cm is used in the conversion region, 3 kV/cm in the transfer regions and 5 kV/cm in the induction region. These values optimize the gain and the electron collection at the anode \cite{Peisert}.\\

\begin{figure}[btp]
		\centering
   		\includegraphics[width=0.4\textwidth]{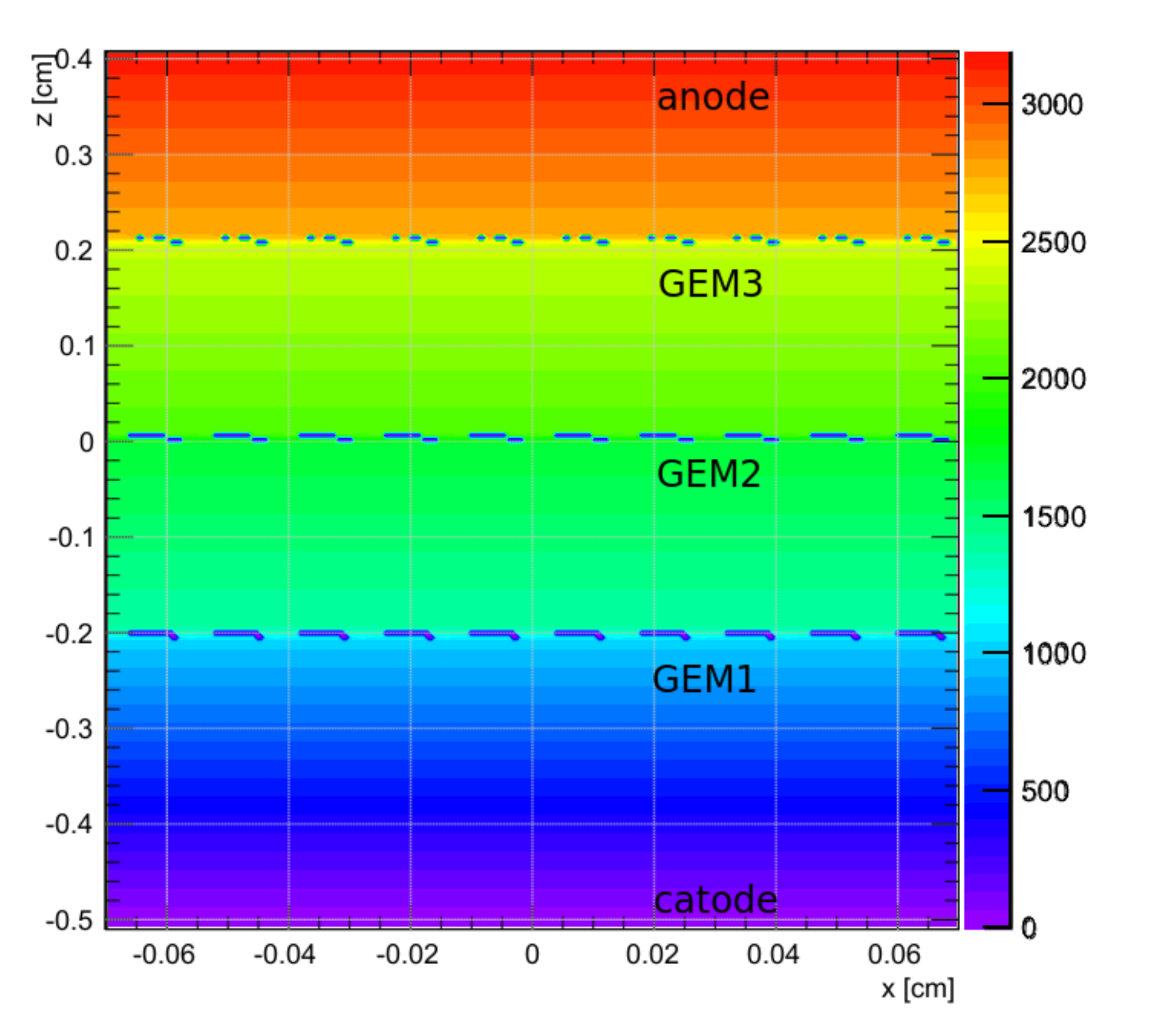}
   		\includegraphics[width=0.38\textwidth]{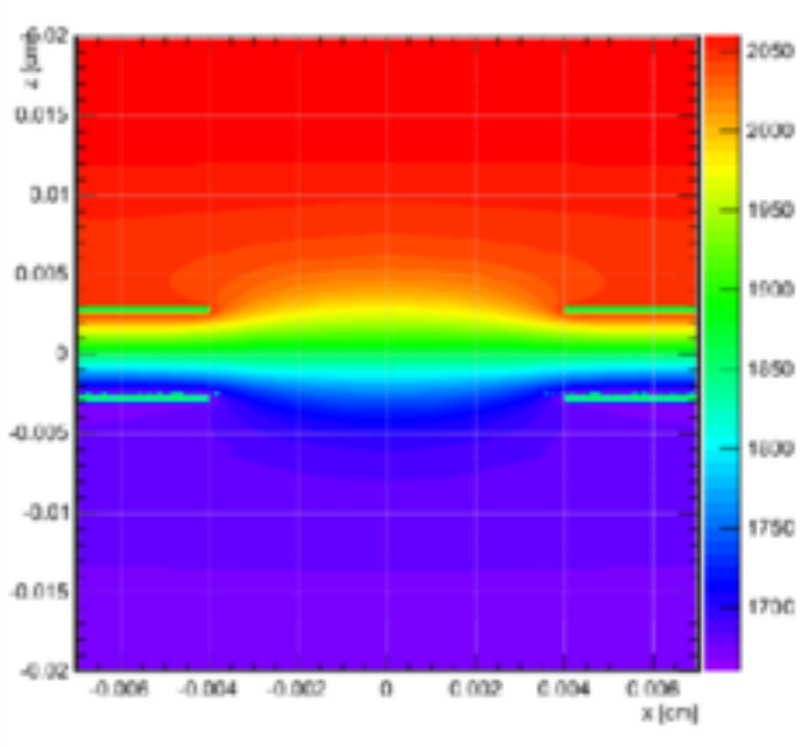}
   		\caption[Garield potential simulations]{Potential used during the simulation in Argon-CO$_2$ in $X-Z$ representation. Different colours show different values. Full triple GEM is shown in (a) and a focus on a single hole in (b).}
	\label{fig:field}
	\end{figure}

Between the two GEM's electrode potential differences of 390/380/370 V (1140 V) are used in order to obtain an applied field of $\sim$ 10$^5$ V/cm in Argon-CO$_2$ (285/275/265 V = 825 V in Argon-CC$_4$H$_{10}$). In this regime electron multiplication takes place. In Fig. $\ref{fig:field}$ and $X-Z$ picture shows the potential for the full triple GEM (on the left) and a single hole (on the right).

\subsection{Simulations of the detector response}

1 GeV/c muons and pions are used to simulate, with Garfield++, the detector response to minimum ionizing particle (m.i.p.); 
the Heed program generates primary electrons along the particle track and the field takes care of the drift to the anode. When the electrons cross the GEM holes, the multiplication occurs. \\

\begin{figure}[htbp]
		\centering
		\begin{subfigure}[h]{0.4\textwidth}
   			\includegraphics[width=\textwidth]{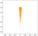}
			\caption{}
			\label{fig:drift:a}
		\end{subfigure}
		\begin{subfigure}[h]{0.415\textwidth}
   			\includegraphics[width=\textwidth]{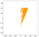}
   			\caption{}
			\label{fig:drift:b}
		\end{subfigure}\\
		\begin{subfigure}[h]{0.42\textwidth}
   			\includegraphics[width=\textwidth]{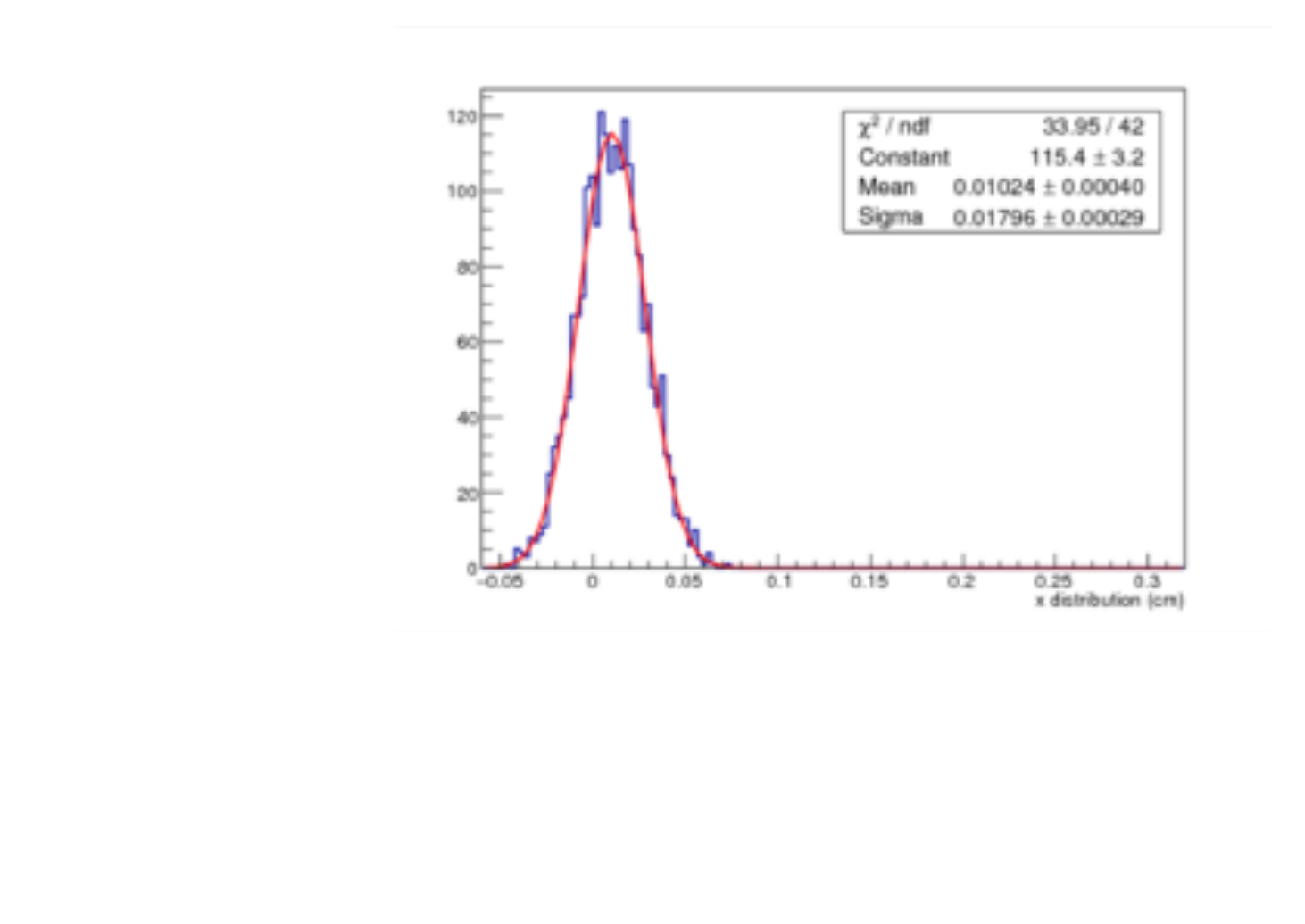}
   			\caption{}
			\label{fig:drift:c}
		\end{subfigure} 
		\begin{subfigure}[h]{0.42\textwidth}
   			\includegraphics[width=\textwidth]{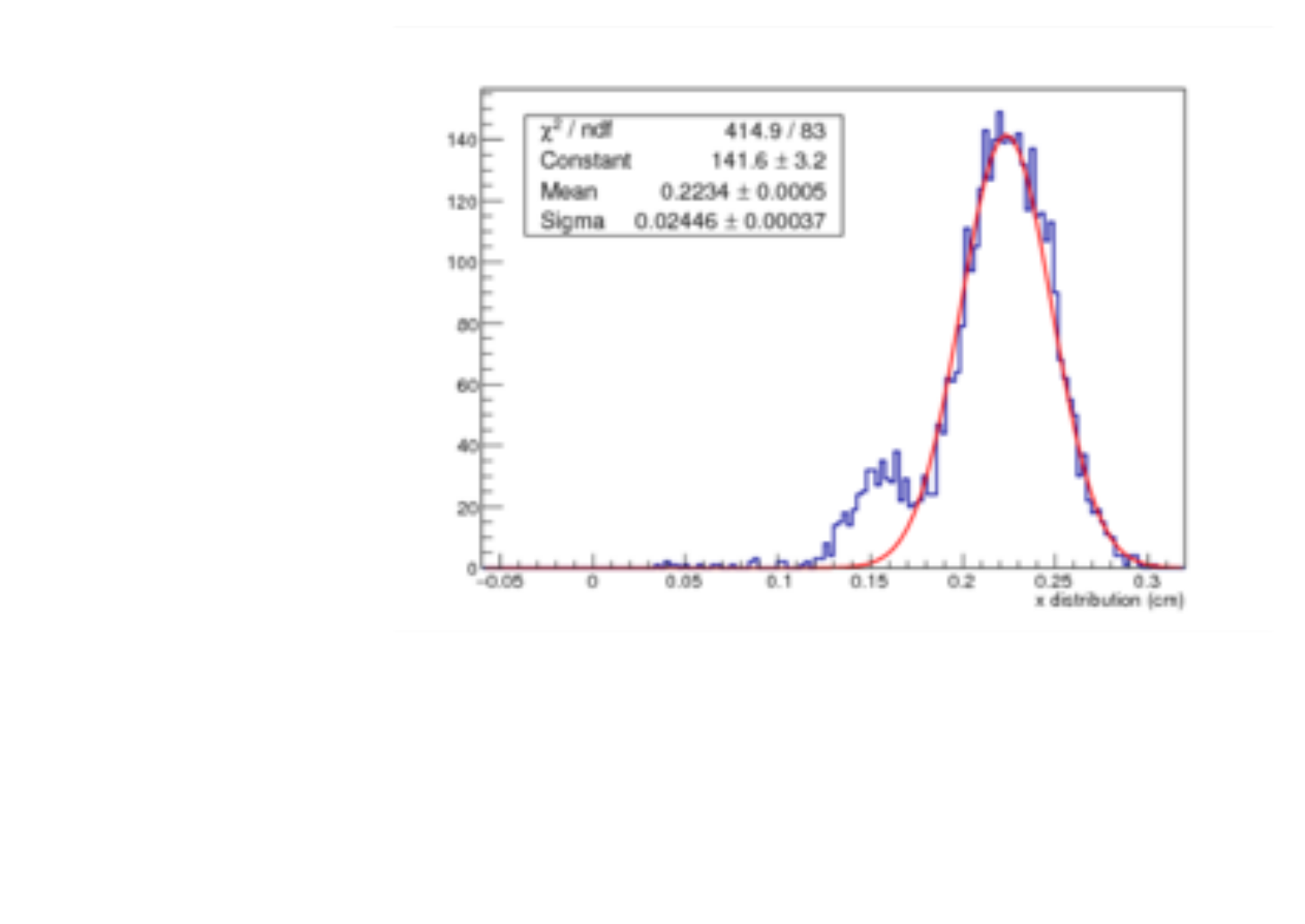}
   			\caption{}
			\label{fig:drift:d}
		\end{subfigure}
	\caption[Drift simulations in Argon-CO$_2$ and 3 mm conversion gap.]{Garfield simulations in Argon-CO$_2$ gas mixture and 3 mm conversion gap. On top a picture of the avalanche drift simulations with (b) and without (a) magnetic field. The bottom plots their charge distribution at anode for B = 0 T (c) and B = 1 T (d). The displacement and the broadening of the electron distribution is can be notice.}
	\label{fig:drift}
	\end{figure}

Fig. \ref{fig:drift} shows the electrons drift and the charge distribution at the anode for Argon-CO$_2$ gas mixture with and without magnetic field.

In Fig. \ref{fig:drift:a} the avalanche propagates parallel to the field direction producing a gaussian charge distribution at the anode. The distribution is approximately centered at zero and the width distribution is about 180 $\mu$m.

Due to the magnetic field effect, in Fig. \ref{fig:drift:b}, the displacement due to the Lorentz force in the avalanche simulations is evident, and the charge distribution is moved with respect to the case B = 0 T by more than 2 mm (see Fig. \ref{fig:drift:d}). The Lorentz angle is measured to be about 15$^{\circ}$ and the shape is still quite gaussian with a tail due to transverse drift.

The simulation framework produced is used to compare the results of different detector configurations as reported in the next section.

\subsection{Analysis of different detector configurations}
\label{sec:gaga}

In addition to the baseline design, 3 mm of conversion gap and Argon-CO$_2$ gas mixture, other configurations have been simulated for a comparison with the beam test results. Such configurations include combinations of a different conversion region of 5 mm and different gas mixtures, such as Argon-C$_4$H$_{10}$.

Increasing the conversion region from 3 mm to 5 mm, a broadening of the electron avalanche is expected at the anode, producing a higher cluster multiplicity but on the other hand increasing the statistical fluctuations of the charge.

\begin{figure}[btp]
		\centering
		\begin{subfigure}[h]{0.38\textwidth}
   			\includegraphics[width=\textwidth]{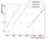}
   			\caption{}
			\label{fig:gasgain:a}
		\end{subfigure}
		\begin{subfigure}[h]{0.42\textwidth}
   			\includegraphics[width=\textwidth]{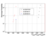}
   			\caption{}
			\label{fig:gasgain:b}
		\end{subfigure}
   		\caption[Gain value and discharge probability for different gas mixtures.]{Gain value and discharge probability for different gas mixtures from experimental data (GDD \cite{gdd}).}
	\label{fig:gasgain}
	\end{figure}

Changing the mixture from Argon-CO$_2$ to Argon-C$_4$H$_{10}$ slightly increases the primary electron, due to the different dE/dx behaviour and allows to decrease the applied voltage reducing the discharge probability, as show in Fig. \ref{fig:gasgain}.

Eight different configurations are studied and the results are summarize in Tab. \ref{tab:garfres}: for each of the eight configurations analyzed the mean center and width of the electron avalanche are extracted from a gaussian fit to charge distribution of the simulated event, (see for example Figs. \ref{fig:drift:c} and \ref{fig:drift:d}). 

The Lorentz angle is calculated from the center value with and without magnetic field and the cluster multiplicity is extracted by means of toy Monte Carlo simulating the readout plane \footnote{A toy Monte Carlo is used to extrapolate the multiplicity information given the electron charge distributions from the Garfield simulation. A guassian distribution is generated with total charge normalized to the charge expected from dE/dx calculation, and the correct strip geometry (pitch).}.

\paragraph{Magnetic field effect.}

As already said, the magnetic field has the effect of broadening and displacing the electron avalanche, increasing the cluster multiplicity. For all the configurations the width of the charge distribution increases from a factor 1.6 to 2.2, and the displacement changes from 2.3 mm to 3.6 mm. As expected the charge distribution along the direction parallel to the magnetic field is unchanged.


\paragraph{Conversion gap effect}

\begin{figure}[btp]
		\centering
   		\includegraphics[width=0.48\textwidth]{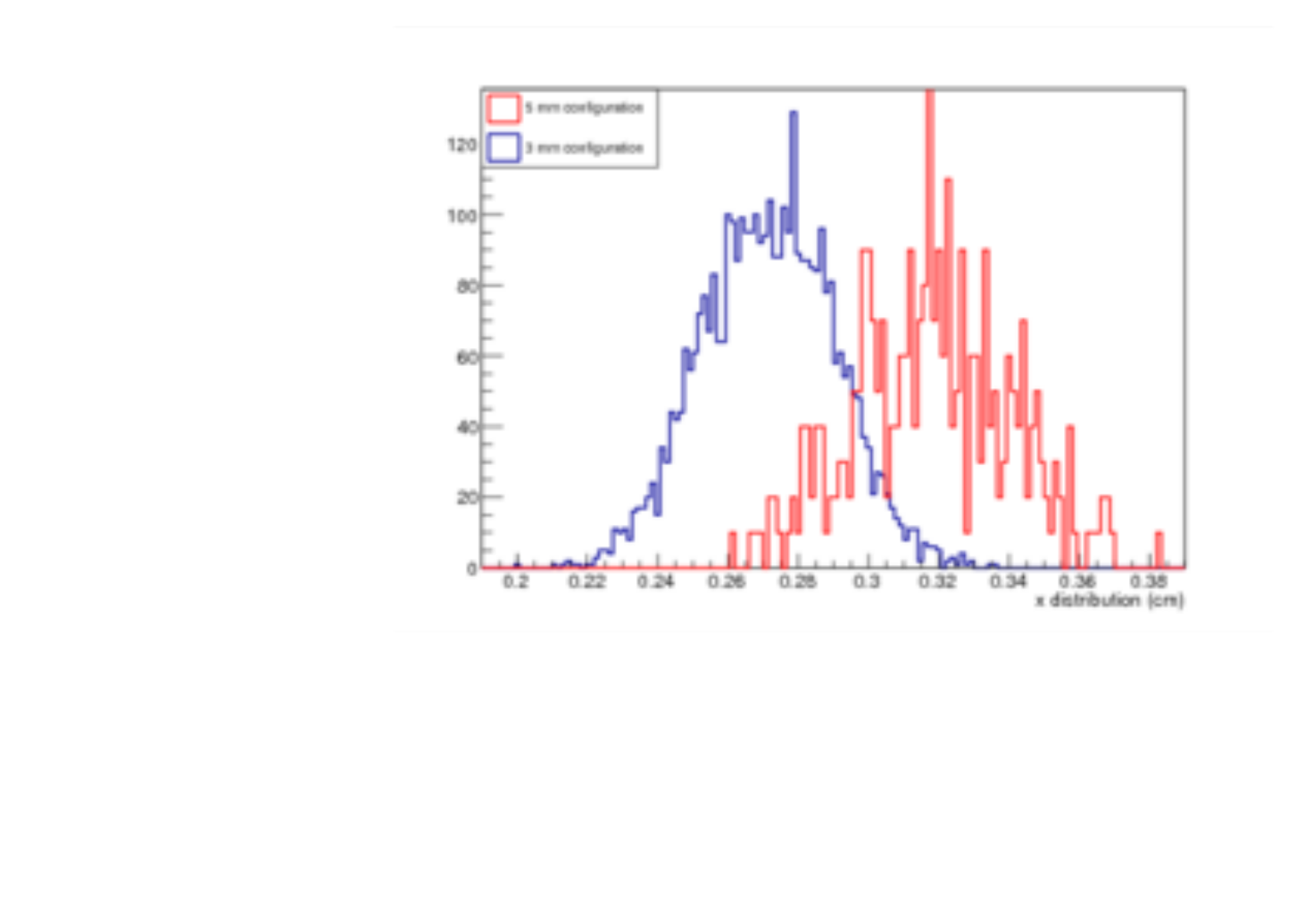}
   		\caption[Charge distribution at anode for different gaps]{Charge distribution at the anode for 3 mm (blue) and 5 mm (red) conversion gap with Argon-C$_4$H$_{10}$ gas mixture.}
	\label{fig:garfgeo}
	\end{figure}

Changing the conversion region from 3 mm to 5 mm a broader electron avalanche is expected. Simulations with different geometries are performed for B = 0 T and B = 1 T and for different gas mixtures. The analysis parameters are shown in Tab. \ref{tab:garfres}. Fig. \ref{fig:garfgeo} reports, as an example, the comparison between the charge distributions at 1 T for one event with the gap of 3 mm and 5 mm. 

The average displacement at 1 T is 370 $\mu$m (Argon-CO$_2$) and 427 $\mu$m (Argon-C$_4$H$_{10}$); due to the larger conversion region the avalanche has 2 mm more to drift in the direction given by the Lorentz angle. The larger effect on the electron avalanche size is for the 5 mm configuration with Argon-C$_4$H$_{10}$ gas mixture which provides a multiplicity of 3.77 .

\paragraph{Gas mixture effect}

	\begin{figure}[btp]
		\centering
		\begin{subfigure}[h]{0.48\textwidth}
   			\includegraphics[width=\textwidth]{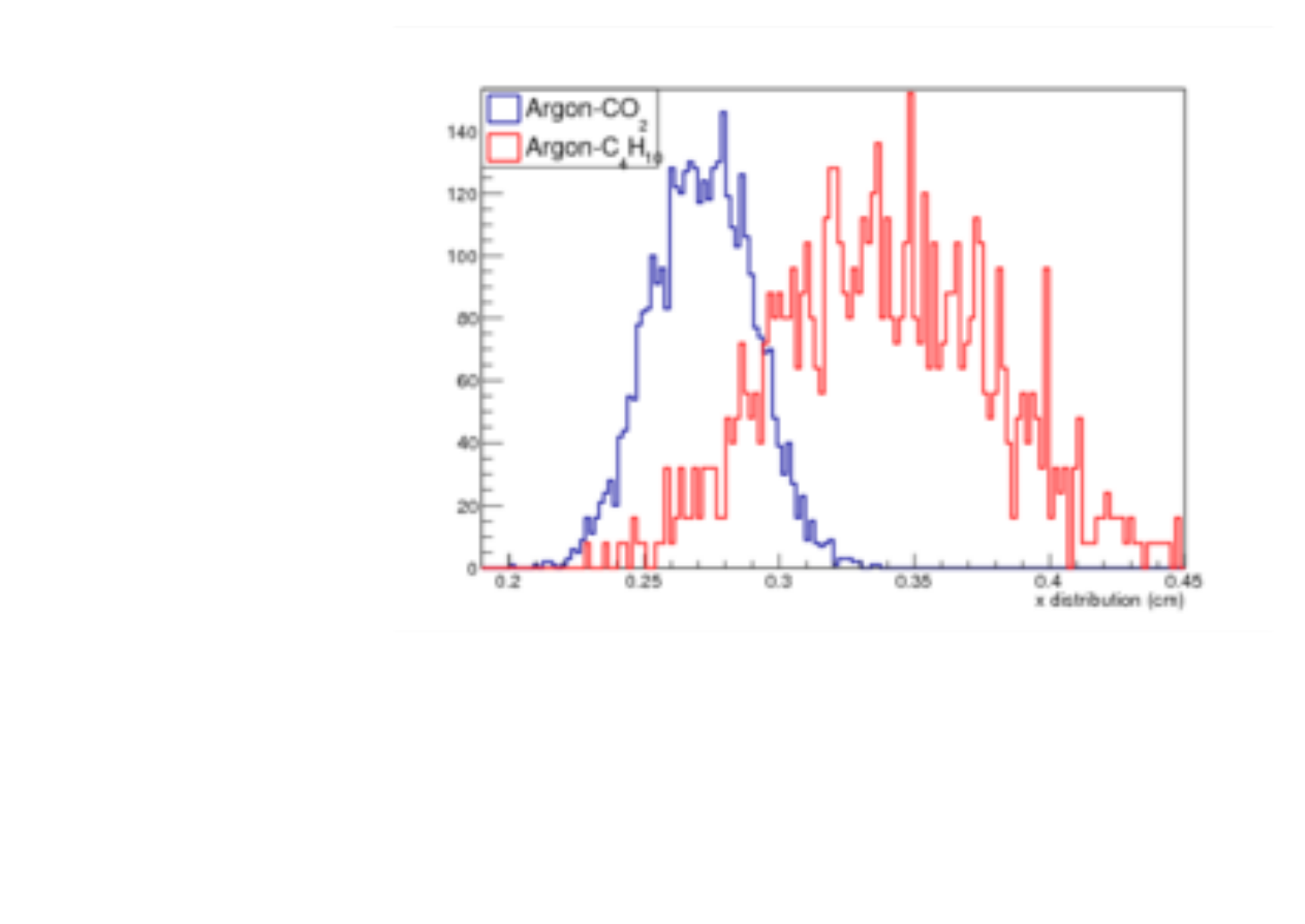}
			\caption{}
			\label{fig:garfgas:a}
		\end{subfigure}
		\begin{subfigure}[h]{0.48\textwidth}
   			\includegraphics[width=\textwidth]{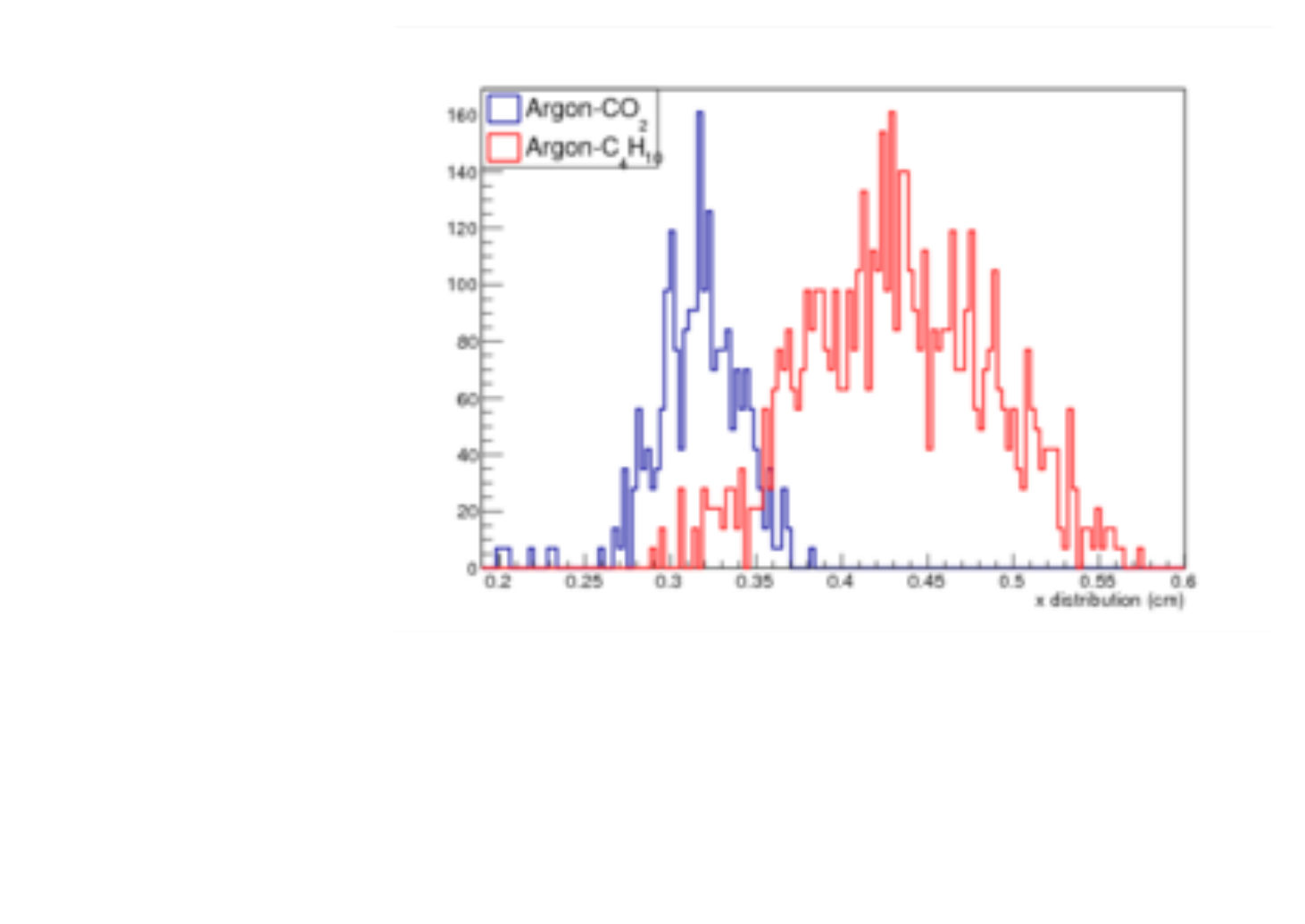}
			\caption{}
			\label{fig:garfgas:b}
		\end{subfigure}
   		\caption[Charge distribution at anode for different gas mixtures]{Charge distribution at the anode for Argon-CO$_2$ (blue) and Argon-C$_4$H$_{10}$ (red) with 3 mm (left) and 5 mm (right) conversion gap.}
	\label{fig:garfgas}
	\end{figure}

Comparing the 3 mm simulation (Fig. \ref{fig:garfgas:a}) at 1 T, the Argon-C$_4$H$_{10}$ displacement is about 800 $\mu$m higher than for the Argon-CO$_2$, and its multiplicity (3.56) is roughly a factor 1.3 larger. The increase is expected by the gas drift properties discussed earlier in this chapter. For the 5 mm simulation (Fig. \ref{fig:garfgas:b}) the increasing of the displacement for Argon-C$_4$H$_{10}$ gas mixture is slightly higher than for the multiplicity.\\

\begin{table}[ht]
\begin{minipage}{0.9\textwidth}
\begin{center}
\begin{tabular}{l c c c c c c} 
\hline
\textbf{Gas}	& \textbf{Conversion}	& \textbf{Magnetic}	& \textbf{$\Delta X$ ($\mu$m)}	& \textbf{$\sigma$ ($\mu$m)}	& \textbf{Lorentz}	& \textbf{Cluster}	\\
		& \textbf{Gap} 		& \textbf{field (T)}	&  				&				& \textbf{angle ($^{\circ}$)} & \textbf{multiplicity}	\\
\hline
Argon-CO$_2$		& 3	& 0	& 38 $\pm$ 72		& 182 $\pm$ 18 	& / &  1.9 \\
\hline
Argon-CO$_2$		& 5	& 0	& -10 $\pm$ 65	 	& 232 $\pm$ 60 	& / &  2.2 \\
\hline
Argon-C$_4$H$_{10}$	& 3	& 0	& -23 $\pm$ 63	 	& 324 $\pm$ 32	& / &  2.6 \\
\hline
Argon-C$_4$H$_{10}$	& 5	& 0	& 20 $\pm$ 75		& 308 $\pm$ 41	& / &  2.5 \\
\hline
Argon-CO$_2$		& 3	& 1	& 2315 $\pm$ 470		& 348 $\pm$ 88	& 14.9	 &  2.7 \\
\hline
Argon-CO$_2$		& 5	& 1	& 2695 $\pm$ 392		& 511 $\pm$ 265	& 14.2	 &  3.6 \\
\hline
Argon-C$_4$H$_{10}$	& 3	& 1	& 3131 $\pm$ 357		& 512 $\pm$ 131	& 20.3	 &  3.6 \\
\hline
Argon-C$_4$H$_{10}$	& 5	& 1	& 3558 $\pm$ 675		& 557 $\pm$ 257	& 18.9	 &  3.8 \\
\hline
\end{tabular}
\caption[Garfield simulation results]{Garfield simulation results for different gas mixtures, magnetic field and conversion gap. Same errors are large due to low statistic.}
\label{tab:garfres}
\end{center}
\end{minipage}
\end{table}

	\chapter{The beam test}
\label{chTesBTeam}

\section{Purpose of the test}

A Beam Test (BT) is a direct experiment where particle beam interacts with a detector prototype to test it in real experimental environment.
A BT has been performed with a planar prototype at SPS H4 beam line at CERN within the RD51 \cite{rd51} collaboration, in December 2014. 

The main purpose of this BT is to measure:
\begin{itemize}
\item efficiency at different gain;
\item cluster size as function of the magnetic field;
\item the signal to noise ratio;
\item spatial resolution as function of the magnetic field;
\item the differences between the gas mixtures Argon-CO$_2$ (70/30) and
Argon-C$_4$H$_{10}$ (90/10).
\end{itemize}

The data acquired during the BT will be used also to valid the Garfield simulation, $e.g.$ the values reported in Tab. \ref{tab:garfres}, in order to tune the output of the Monte Carlo with real data to allow the simulation to explore the widest range of possible configurations. The output of the BT studies together with the output of the Garfield simulations will be used to improve the reliability of the full GEANT4 simulation of the detector.

Another goal of the BT is the measurement of the performance of the analog readout in the magnetic field and the validation of the BESIII anode configuration.

\begin{figure}[btp]
		\centering
   		\includegraphics[width=0.6\textwidth]{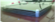}
   		\caption[Picture of the planar prototype.]{Picture of the planar prototype without the protection to keep inside the gas mixture.}
	\label{fig:proto}
	\end{figure}

We tested a planar triple-GEM prototype with two different gap sizes (3 and 5 mm) and two gas mixtures (Argon-CO$_2$ and Argon-C$_4$H$_{10}$). Currently we analyzed only the data for the Argon-C$_4$H$_{10}$ and 5 mm configuration. A picture is reported in Fig. \ref{fig:proto}.

\section{The experimental setup}

The H4 beam line at north area beam facility at the SPS (CERN) has a secondary extracted beam of pions and muons with momentum up to 400 GeV/c. The facility is provided of a dipole magnet (GOLIATH) capable to reaches magnetic field of 1.5 T.

Our experimental setup is consists of:
\begin{itemize}
\item a trigger system;
\item a tracking telescope;
\item the BESIII prototype;
\item the gas and HV system.
\end{itemize}

\begin{figure}[btp]
		\centering
   			\includegraphics[width=0.8\textwidth]{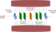}
   		\caption[Test Beam setup]{Drawing of the BT setup, the particles go from left to right. The BESIII test chamber (yellow) is in the middle around 4 tracker (green), 2 forward and 2 backward, and 4 triggers (blue), 2 each cross, by side. The full setup is placed in a magnet (violet).}
	\label{fig:BTsetup}
	\end{figure}

A schematic drawing of the setup is shown in Fig. \ref{fig:BTsetup}.

The trigger system is made by four scintillator bars readout by Silicon PhotoMultipliers (SiPM) through a wave-length shifting fiber glued inside the scintillator. The modules have been assembled and characterized in Ferrara showing a detection efficiency higher than 95\%.\\

The tracking system is composed by four planar 10$\times$10 cm$^2$ triple-GEM with 3/2/2/2 gap geometry (cathode/transfer1/transfer2/anode) and orthogonal $XY$ strip readout. The pitch of the strips of the tracker chambers is 650 $\mu$m providing a position measurement with a precision of about 45 $\mu$m. \\

The BESIII prototype is also a 10$\times$10 cm$^2$ triple GEM with orthogonal $XY$ strips. The pitch of the strips is 650 $\mu$m and the width is 570 $\mu$m for the $x$ coordinate and 130 $\mu$m for the $y$ as it will be in the final detector. The ground plane is at 2 mm from the anode. The conversion gap of the prototype has been changed from 3 mm to 5 mm during a short shutdown of the BT.\\

Both the tracking chambers and the BESIII prototype have been acquired by the Scalable Readout System (SRS) developed within the RD51 Collaboration for the readout of Micro Pattern Gas Detector \cite{srs}. The SRS  is split in three parts. In the front-end of the system, close to the detector, a front-end hybrid carries the front-end ASIC (APV25) together with all necessary supporting circuits (discharge protection, power regulators, etc.). On the other end of the DAQ unit there is a FPGA-based card called
Front-End Card (FEC), which contains most complex circuits of the DAQ unit (programmable logic, memory, high-speed communication).\\ 

The APV25 is an analog sampling chip with 128-channels. Each channel of the APV25 is a charge-sensitive preamplifier. The amplifier output amplitudes are sampled and stored in a cell. When an external trigger arrives at the chip, the cells corresponding to the known trigger latency are flagged for readout. The multiplexed analog data stream from each APV25 chip is digitized by a 10 bit flash ADC. 27 samples are readout for each event, the sampling time is 25 ns.\\

The HV is provided by a commercial CAEN system instrumented with CAEN A1550 power supplies \cite{caen} and distributed to the GEM electrodes by a custom distribution system. Singular current monitoring is provided by a nano amperometer. The gas is supplied by premixed bottles and it flows through the chambers, connected in series.\\

\begin{figure}[btp]
		\centering
		\begin{subfigure}[h]{0.62\textwidth}
   			\includegraphics[width=\textwidth]{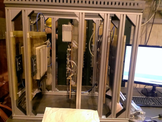}
			\caption{}
			\label{fig:pictureofyou:a}
		\end{subfigure}
		\begin{subfigure}[h]{0.35\textwidth}
   			\includegraphics[width=\textwidth]{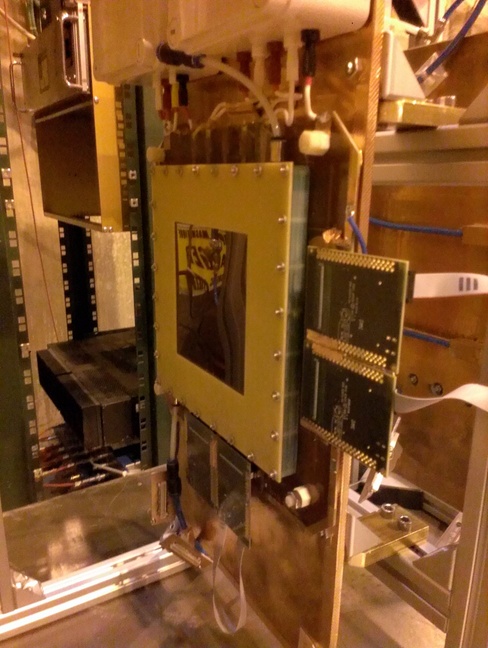}
			\caption{}
			\label{fig:pictureofyou:b}
		\end{subfigure}\\
		\begin{subfigure}[h]{0.98\textwidth}
   			\includegraphics[width=\textwidth]{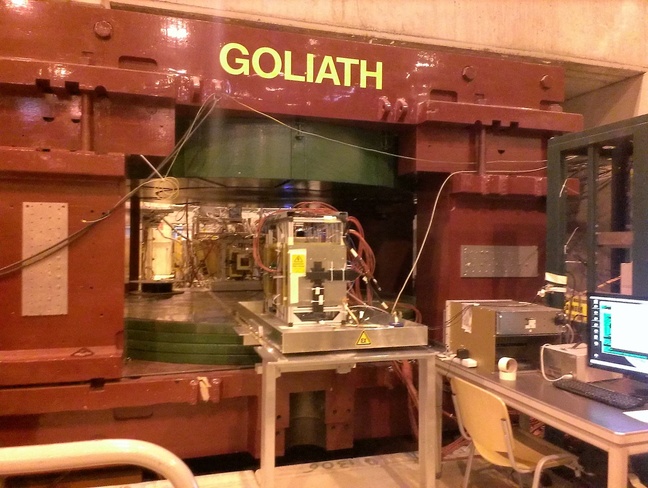}
			\caption{}
			\label{fig:pictureofyou:b}
		\end{subfigure}
   		\caption[Pictures of the setup.]{Pictures of the setup: (a) shows the tracking telescope and the BESIII test chamber in the middle; (b) focuses the BESIII test chamber (c) is a picture of the full setup of the beam test inside the magnet.}
	 	\label{fig:pictureofyou}
	\end{figure}

Pictures of the setup are shown in Fig. \ref{fig:pictureofyou}.

\section{The data taking}

During the 3 weeks of run the following data have been acquired:
\begin{itemize}
\item an HV scan with a gain ranging from 2k - 22k with Argon-C$_4$H$_{10}$ and Argon-CO$_2$ gas mixtures; 
\item a magnetic field scan from -1 T to 1 T for both gas mixtures;
\item different incident angles (0$^{\circ}$/10$^{\circ}$/30$^{\circ}$/45$^{\circ}$) in Argon-C$_4$H$_{10}$ with  magnetic field of -1/0/1 T.
\end{itemize} 

Up to now only Argon-C$_4$H$_{10}$ have been studied. The analyzed data are summarized in Tab. \ref{tab:datataken}. 

\begin{table}[ht]
\begin{minipage}{0.9\textwidth}
\begin{center}
\begin{tabular}{ccccc}
\hline
\textbf{Gain} & \textbf{B = 0 T}&  \textbf{B = -0.5 T}	&  \textbf{B = -1 T}	&  \textbf{B = 1 T} \\
\hline
800 V	& 24 k	& x	& x 	& x 	\\
\hline
2000	& 21 k	& 15 k	& 22 k 	& 21.5 k 	\\
\hline
3000	& 16.2 k	& 20 k	& 24 k 	& 20 k 	\\
\hline
4500	& 20 k	& 15 k	& 20 k 	& 21 k 	\\
\hline
6500	& 5.5 k	& 15 k	& 20 k 	& 20 k 	\\
\hline
10000	& 48 k	& 20 k	& 20 k 	& 21 k 	\\
\hline
16000 	& 24 k	& 15 k	& 20 k 	& 20 k 	\\
\hline
20000 	& 15 k	& 20 k	& 19 k 	& 20 k 	\\
\hline

\end{tabular}
\caption[Summary of Argon-C$_4$H$_{10}$ data analyzed.]{Summary Argon-C$_4$H$_{10}$ data analyzed.}
\label{tab:datataken}
\end{center}
\end{minipage}
\end{table}


\section{The reconstruction software}
\label{sec:algorith}

The raw data are processed by an offline software code developed by INFN-Ferrara. 
The event output is formed by 27 charge value (corresponding to the 27 time samples) for each channel.\\

A first algorithm digitizes the charge signal and maps it to a physical strip ($hit$). Then the clusterization occurs, adjacent strips are grouped in the same 1-D $cluster$. The position of the cluster is calculated through the charge centroid as described in Sec. \ref{sec:anal}. 
A $cluster$ in the $x$ view and a $cluster$ in the $y$ view form a 2-D $cluster$. A 2-D $cluster$ contains 3-D position, charge and time information of the electron avalanche.\\

\begin{figure}[btp]
		\centering
   		\includegraphics[width=0.6\textwidth]{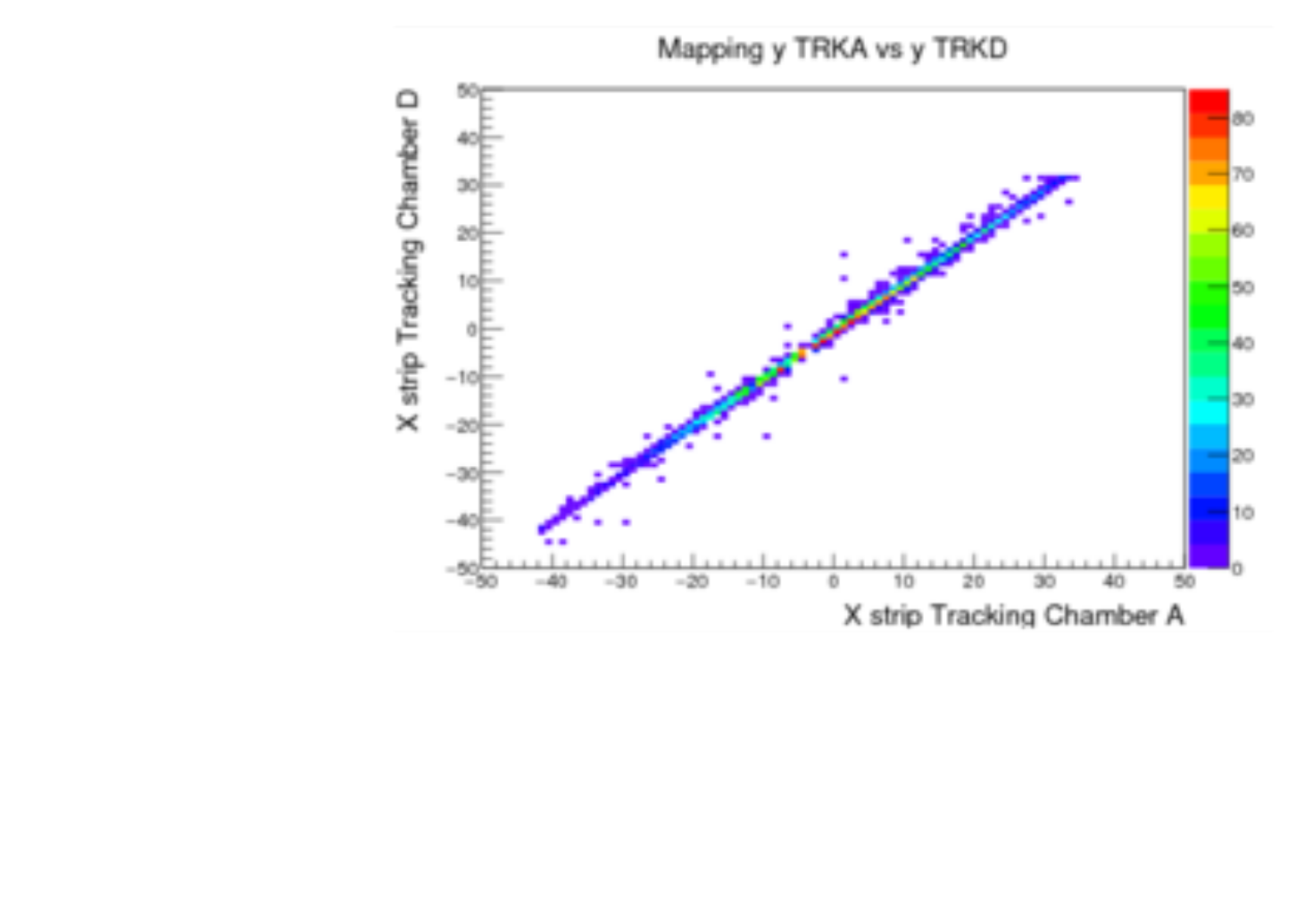}
   		\caption[Hit-map produced at the beam test.]{Hit-map produced at the beam test of the $x$ view for two tracking chambers	.}
	\label{fig:hitmap}
	\end{figure}

\begin{figure}[btp]
		\centering
		\begin{subfigure}[h]{0.55\textwidth}
   			\includegraphics[width=\textwidth]{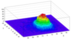}
			\caption{}
			\label{fig:BTbeam:a}
		\end{subfigure}
		\begin{subfigure}[h]{0.44\textwidth}
   			\includegraphics[width=\textwidth]{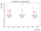}
			\caption{}
			\label{fig:BTbeam:b}
		\end{subfigure}
   		\caption[BESIII test chamber beam profile and $x$ view of a single track.]{a) BESIII test chamber beam profile of a single run. b) $x$ view of a single track without alignment.}
	\label{fig:BTbeam}
	\end{figure}

After checking the goodness of the cabling with hit-map plots (see Fig. \ref{fig:hitmap}) the beam profile and straight tracks could be reconstructed, Fig. \ref{fig:BTbeam}.

If all the clusters of a chamber from the same run are plotter together, it is possible to see the beam profile as shown in Fig. \ref{fig:BTbeam:a}.

\section{Data analysis}

\subsection{Noise suppression}

\begin{figure}[btp]
		\centering
		\begin{subfigure}[h]{0.48\textwidth}
   			\includegraphics[width=\textwidth]{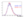}
			\caption{}
			\label{fig:BTnoise:a}
		\end{subfigure}
		\begin{subfigure}[h]{0.48\textwidth}
   			\includegraphics[width=\textwidth]{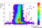}
			\caption{}
			\label{fig:BTnoise:b}
		\end{subfigure}
   		\caption[Pedestal charge distribution and charge distribution as function of the sampling time.]{a)Pedestal charge distribution for the same channel with B = 0 T (blue) and B = 1 T (red). The scale is arbitrary. b) Charge distribution as function of the sampling time. Red lines show the applied cut.}
	\label{fig:BTnoise}
	\end{figure}

Pedestal runs are used to subtract the electronic noise. Fig. \ref{fig:BTnoise:a} shows an example of typical pedestal with and without magnetic field. The magnetic field operation does not introduce any additional noise. Pedestal runs have been acquired every 8 hours to take into account changes in the noise level. The data acquisition provides automatic pedestal subtraction; 80\% of the pedestal is removed online. The charge distribution as function of the sampling time for muon events has been used to study additional cuts on the charge and on the sampling time to remove the remaining background. The signal is concentrated between the samples 5 and 20. The hits outside  these boundaries are considered noise and are used to study the cut on the strip charge ($>$ 50 ADC channels).

\subsection{Data without magnetic field}
\label{sec:BT0}

Events with one 2-D $cluster$ for each tracking chamber are selected. Tracks are reconstructed from the clusters in the tracking chamber. A fit is performed to the track clusters and the residual distributions are used to align the detector. After the alignment we look for clusters in the BESIII chamber in order to extract efficiency, cluster size and resolution. Only clusters associated to the tracks are considered.

\paragraph*{Efficiency} 
\label{sec:BTeff}

	\begin{figure}[btp]
		\centering
		\begin{subfigure}[h]{0.48\textwidth}
   			\includegraphics[width=\textwidth]{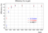}
			\caption{}
			\label{fig:BTeff0:a}
		\end{subfigure}
		\begin{subfigure}[h]{0.48\textwidth}
   			\includegraphics[width=\textwidth]{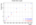}
			\caption{}
			\label{fig:BTeff0:b}
		\end{subfigure}

   		\caption[Efficiency and cluster size of the BESIII test chamber as function of the gain.]{Efficiency (a) and cluster size (b) of the BESIII test chamber as function of the gain. Blue dot represent the cluster along the $x$ coordinate, red along the $y$ coordinate.}
	\label{fig:BTeff0}
	\end{figure}

The efficiency of the BESIII test chamber is calculated at different gain values reported in Tab. \ref{tab:datataken}. Fig. \ref{fig:BTeff0:a} shows the efficiency for the $XY$ view separately and for the coincidence as function of the gain. The plateau starts about a gain of 6500. In the plateau region the average efficiency for the 2-D $cluster$ is about 97 \%.
The working point is around a gain of 10000 (820 V), see Fig. \ref{fig:BTeff0:a}, over this value the efficiency shows a saturation effect.

\paragraph*{Cluster size} 
Cluster information is extracted for the BESIII chamber. Results are reported in Fig. \ref{fig:BTeff0:b} as function of the gain; the cluster multiplicity ranges from 1.9 to 3.9 for the $Y$ clusters and from 2 to 4 for $X$ clusters, showing a non equal charge sharing. The behaviour is linear at low gain while at higher gains the linearity is lost due to saturation of the amplification. 

\paragraph*{Resolution} 

\begin{figure}[btp]
		\centering
		\begin{subfigure}[h]{0.48\textwidth}
   			\includegraphics[width=\textwidth]{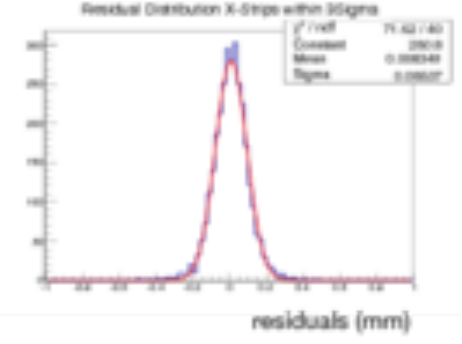}
			\caption{}
			\label{fig:BTresidual:a}
		\end{subfigure}
		\begin{subfigure}[h]{0.48\textwidth}
   			\includegraphics[width=\textwidth]{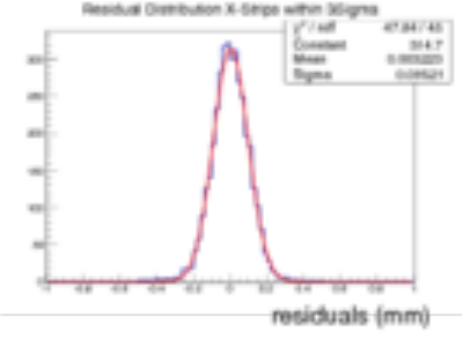}
			\caption{}
			\label{fig:BTresidual:b}
		\end{subfigure}

   		\caption[BESIII test chamber residual distribution.]{BESIII test chamber residual distribution for a gain of 10k (a) and 20k (b).}
	\label{fig:BTresidual}
	\end{figure}

\begin{figure}[btp]
		\centering
   		\includegraphics[width=0.8\textwidth]{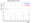}
   		\caption[BESIII test chamber resolution.]{BESIII test chamber resolution ($\mu$m) for $x$ and $y$ view as function of the gain.}
	\label{fig:BTres0}
	\end{figure}

The resolution without magnetic field is shown in Fig. \ref{fig:BTres0}. The resolution is measured as the width of the gaussian fit to the residual distribution ($e.g.$ see Fig. \ref{fig:BTresidual}). In the plateau region the average resolution is about 90 $\mu$m that became about 80 $\mu$m deconvolving the resolution of the tracking system.

\subsection{Data with magnetic field}
\label{sec:BT1}

Only preliminary results from the data with magnetic field will be presented since data analysis is still on going and it is beyond of the scope of this work. Only the $y$ coordinate will be affected by the magnetic field.

\paragraph*{Efficiency}

	\begin{figure}[btp]
		\centering
   		\includegraphics[width=0.8\textwidth]{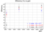}
   		\caption[Efficiency of the BESIII test chamber with magnetic field as function of the gain.]{Efficiency of the BESIII test chamber with magnetic field as function of the gain.}
	\label{fig:BTeff1:a}
	\end{figure}

	\begin{figure}[btp]
		\centering
		\begin{subfigure}[h]{0.48\textwidth}
   			\includegraphics[width=\textwidth]{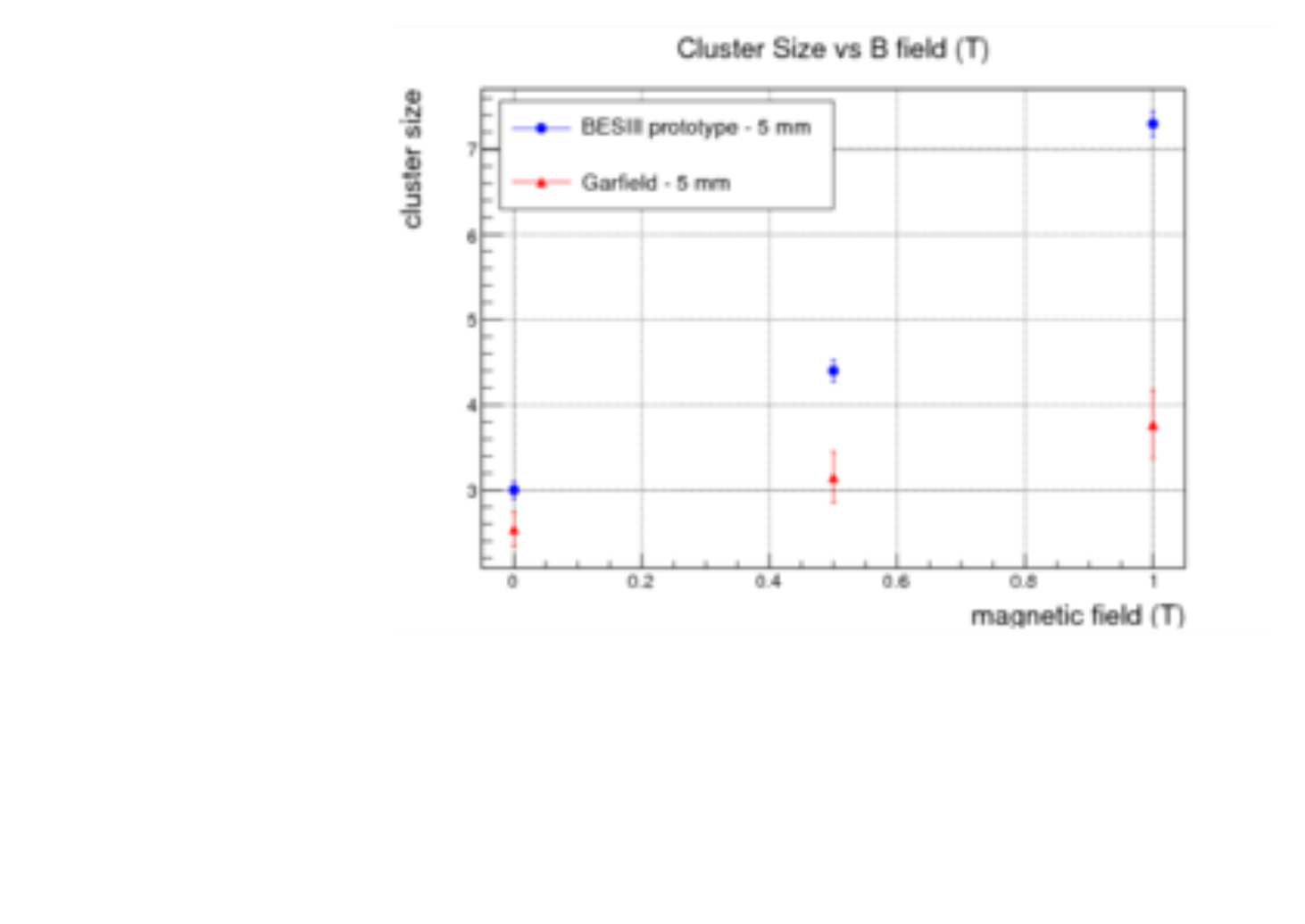}
			\caption{}
			\label{fig:BTclcl1:a}
		\end{subfigure}
		\begin{subfigure}[h]{0.48\textwidth}
   			\includegraphics[width=\textwidth]{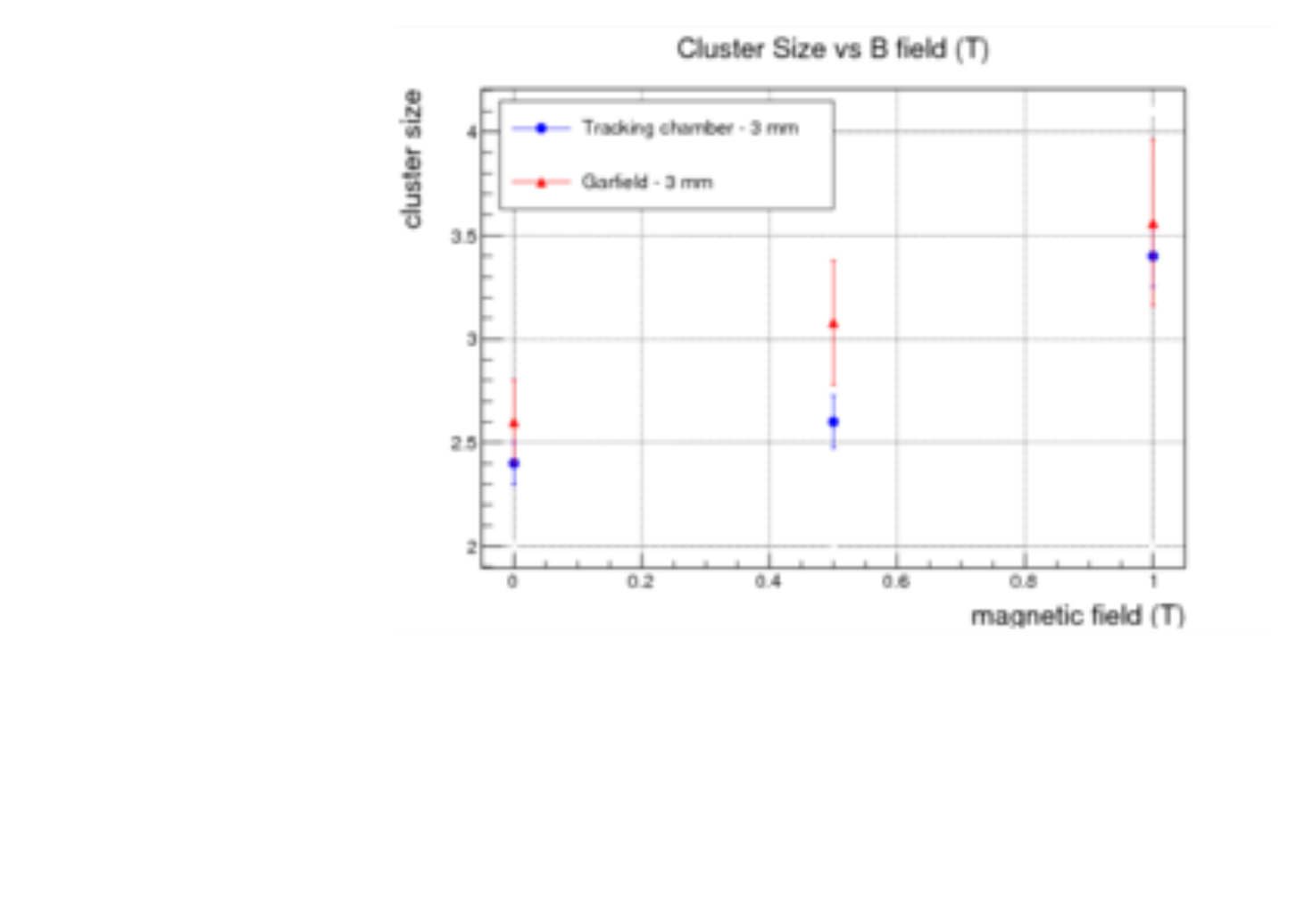}
			\caption{}
			\label{fig:BTclcl1:b}
		\end{subfigure}
		\caption[Cluster size with magnetic field for real and simulated data in 3 mm and 5 mm configurations.]{Cluster size with magnetic field for real and simulated data in 3 mm and 5 mm configurations.}
	\label{fig:BTclcl1}
	\end{figure}

	\begin{figure}[btp]
		\centering
		\begin{subfigure}[h]{0.48\textwidth}
   			\includegraphics[width=\textwidth]{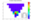}
			\caption{}
			\label{fig:BTeff1:c}
		\end{subfigure}
		\begin{subfigure}[h]{0.48\textwidth}
   			\includegraphics[width=\textwidth]{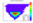}
			\caption{}
			\label{fig:BTeff1:d}
		\end{subfigure}
		\caption[Cluster distribution as function of the cluster size with and without magnetic field.]{Cluster distribution as function of the cluster size with (b) and without magnetic field (a).}
	\label{fig:BTcl1}
	\end{figure}

Using the same procedure described earlier in this chapter, the efficiency with magnetic field (0.5 T and 1 T)is extracted and shown in Fig. \ref{fig:BTeff1:a} as function of the gain. The results are compatible with the ones without the magnetic field.

\paragraph*{Cluster size} 

The effect of the magnetic field is sizable looking at the cluster size; the magnetic field drifts the electron avalanche in the direction of the Lorentz force and broads the charge distribution. As a result the cluster size increases linearly with the magnetic field, see Fig. \ref{fig:BTclcl1:a} for data at gain of 10000.

The same plot reports also the cluster size values from the Garfield simulation: the disagreeament is clear both for the absolute values and for the different slope. Checking the HV parameters for the BESIII chamber with the 5 mm gap, we noticed that the electric field in the conversion gap has been set to a lower value (0.9 kV/cm instead of 1.5 kV/cm) increasing the Lorentz angle and broadening the charge distribution. As a double check we extract cluster size of the tracking chamber (3 mm gap) and the comparison with the Garfield simulation reported in Fig. \ref{fig:BTclcl1:b} shows a good agreement.

In Fig. \ref{fig:BTeff1:c} and \ref{fig:BTeff1:d} the charge profile for $x$ and $y$ clusters is reported as function of the cluster size. The $x$ cluster multiplicity is much lower since it is unaffected by the magnetic field. On the $y$ coordinate, the multiplicity is not only higher but the charge profile departs from the gaussian shape. This can be due to the wrong HV setting just mentioned.

\subsection{Summary of the results}

For the BT results produced so far, we can draw the following conclusions:
\begin{itemize}
\item the efficiency with and without the magnetic field is about 97\% and it reaches the working point at about a gain of 10$^4$;
\item the cluster size increases linearly with the gain and the magnetic field, as expected;
\item the spatial resolution without magnetic field reaches a value of about 80 $\mu$m at a gain of 6500. This result is compatible with the COMPASS resolution reported in Sec. \ref{art}.
\end{itemize}

	\chapter{Conclusions and outlook}

A new inner tracker has been proposed as a replacement of the BESIII experiment inner drift chamber that is suffering from early ageing due to the increasing of the machine luminosity and some high background during the first years of operation. The new inner tracker will be composed by three layers of cylindrical triple GEM detectors.
This work provides input for the detector optimization and for the GEANT4 full simulation that will be used to extract the CGEM-IT expected performance. \\

\noindent
A study of the expected background on the new inner tracker has been performed combining real background events and Monte Carlo data. The CGEM-IT occupancy has been calculated and a maximum rate of 60 kHz per strip is predicted. This result will be used to optimize the design of the front-end electronics and to increase the reliability of the GEANT4 simulation. \\

\noindent
The shape of the electron avalanche has been studied by means of a Garfield simulation. The charge distribution width and the cluster size have been compared for different detector configurations and gas mixtures. From the comparisons between simulations with and without magnetic field the Lorentz angle has been calculated. Garfield simulations, once validated, will be used to extend the beam test studies. \\ 

\noindent
A beam test has been performed in December 2014 with a muon beam to test the GEM analog readout in magnetic field and 
to validate the Garfield simulation and extract input for the digitization. Several configurations have been tested with and without magnetic field:
\begin{itemize}
\item different gap size (3 mm and 5 mm);
\item different gas mixture Argon-CO$_2$ and Argon-C$_4$H$_{10}$;
\item different gas gains.
\end{itemize}

From the results of the analysis, we can draw the following conclusions:
\begin{itemize}
\item the efficiency with and without the magnetic field is about 97\% and it reachs the working point at about a gain of 10$^4$;
\item the cluster size increases linearly with the gain and the magnetic field, as expected;
\item the spatial resolution without magnetic field reachs a value of about 80 $\mu$m at a gain of 6500. This result is compatible with the COMPASS resolution reported in Sec. \ref{art}.
\end{itemize}

\noindent
With the work of this thesis we provide also a set of software and analysis tools that can be used to complete and extend the studies done so far. For the Garfield
simulation more accurate measurements can be done both by increasing the statistics, and improving the output of the simulation trying to reproduce the
gain of the GEM chambers.
The beam test analysis is also under way and needs to be completed. The 3 mm configuration and the Argon-CO$_2$ gas mixture must be fully analyzed, and a new
beam test is foreseen for the end of May 2015 in order to increase the statistics and to take data with different HV settings in the conversion gap that caused the
distortion of the electron distributions with high magnetic field.
Additional innovative studies can be performed with the data acquired so far: a $\mu TPC$ mode readout, that combine time and charge information, can be explored
to have better precision of the cluster position.

\chapter*{Acknowledgements}
I would like to thank Dott. Gianluigi Cibinetto, Gigi, because during this year has been a mentor, which shared with me his passion for this job, and this allowed me to sink into in with pleasure.
To work together has been very formative, both in office and by night in random Control Room  around the world.\\

The application of several years of studies in this project, the CGEM, gave to me great satisfaction during each little result reached.
Thanks to Diego, he has always supported the work done and the cohesion of the research group. It is enjoyable to be  in a unified group. It creates an excellent environment. 
Thanks to Prof. Mauro Savrié, with his method, during the days passed together in the laboratory taught to me a lot.
This year of this thesis showed to me what really is the Research and what is the pleasure in reaching each aim.\\

Thanks to Giorgio for the nice night debate about physics, close the limit of reality, pleasant. To my every-day friends, they stay always close to me. To Serena.\\

A special acknowledgement goes to my family, landmark of the adventure of my life: my parents and my bro Gianluca.
\\\\
Rise up this mornin', \\
Smiled with the risin' sun, \\
Three little birds \\
Pitch by my doorstep \\
Singin' sweet songs \\
Of melodies pure and true, \\
Sayin'...\\

	\chapter*{}

\footnote{The game.}

\end{document}